\documentclass[superscriptaddress,showpacs,amsmath,amssymb]{revtex4}
\usepackage{graphicx}
\usepackage{color}
\usepackage{dcolumn}
\usepackage{mathtools}
\usepackage{ulem}
\usepackage{color}

\usepackage{amsmath}
\usepackage{amssymb}
\usepackage{graphicx}
\usepackage{accents}
\usepackage{pstricks,pst-node,pst-coil,pst-plot}
\usepackage{mathtools}
\usepackage{ulem}
\usepackage{mathtools}
\usepackage{ulem}
\usepackage{graphicx}
\usepackage{color}
\usepackage{graphicx}
\usepackage{color}
\usepackage{dcolumn}
\usepackage{graphicx}
\usepackage{color}
\usepackage{dcolumn}
\usepackage{bm}
\usepackage{mathrsfs}
\usepackage{graphicx}
\usepackage{color}
\usepackage{dcolumn}
\usepackage{bm}
\usepackage{slashed}
\usepackage{epstopdf}
\usepackage{slashed}
\usepackage{bm}
\usepackage{slashed}
\usepackage{dcolumn}
\usepackage{bm}
\usepackage{slashed}
\usepackage{mathrsfs}
\usepackage{graphicx}
\usepackage{color}
\usepackage{bm}
\usepackage{slashed}
\usepackage{epstopdf}
\usepackage{bm}
\usepackage{slashed}
\newcommand{\be}{\begin{equation}}
\newcommand{\ee}{\end{equation}}

\def\d{\mathrm{d}}

\usepackage{amsmath}

\numberwithin{equation}{section}

\begin{document}
\title{Bondi mass, memory effect and balance law of polyhomogeneous spacetime}

\author{Xiaokai He}
\email{sjyhexiaokai@hnfnu.edu.cn}
 \affiliation{School of Mathematics and
Statistics, Hunan First Normal University, Changsha
410205, China}

\author{Xiaoning Wu}
\email{wuxn@amss.ac.cn}
\affiliation{Institute of Mathematics, Academy of Mathematics and Systems Science and State Key Laboratory of Mathematical Sciences,
	Chinese Academy of Sciences, Beijing 100190, China}

\author{Naqing Xie}
\email{nqxie@fudan.edu.cn}
 \affiliation{School of Mathematical Sciences, Fudan
University, Shanghai 200433, China}

\begin{abstract}
Spacetimes with metrics admitting an expansion in terms of a combination of powers of $1/r$ and $\ln r$ are known as polyhomogeneous spacetimes. The asymptotic behaviour of the Newman-Penrose quantities for the vacuum polyhomogeneous spacetimes is presented under certain gauges. The Bondi mass is revisited via the Iyer-Wald formalism. The memory effect of the gravitational radiation in the vacuum polyhomogeneous spacetimes is also discussed. It is found that the appearance of the logarithmic  terms does not affect the balance law and  it remains identical to that of spacetimes with metrics admitting an expansion in terms of powers of $1/r$.

\end{abstract}

\maketitle



\section{Introduction}\label{S1}

Over a century ago, Einstein predicted gravitational waves based on an analysis of the linearised field equations of general relativity \cite{Einstein1916}. However, due to the complexity of diffeomorphisms, the existence of gravitational waves caused a huge controversy for quite a long time \cite{Pirani1956,Ken07}.

Bondi {\it et al.} solved this debate theoretically by proposing the
 Bondi-Sachs (BS) framework \cite{Bondi62,Sachs62}.  The asymptotic behaviour of metric functions in the BS coordinate system is obtained by solving the Einstein field equations using the formal series expansions in powers of $1/r$, where the coordinate $r$ denotes the luminosity distance. Within the BS framework,  the Bondi mass was successfully defined and it satisfied the well-known mass loss formula \cite{Bondi62,Sachs62}. The Bondi-Metzner-Sachs (BMS) group is the asymptotic symmetry group and the Bondi mass possesses a uniqueness property \cite{Chru98}. Newman, Penrose and Unti also used
 the formal $1/r$ expansions with $r$ the affine parameter of a future-oriented null geodesic to study the asymptotically flat spacetimes within the Newman-Penrose (NP) formalism \cite{NP62, NU62, NP65, NP68}.  Bondi {\it et al.} excluded the logarithmic terms in the asymptotic expansions by precluding the appearance of $1/r^2$ terms in some of  the metric functions.This assumption is known as the outgoing radiation condition.  Newman {\it et al.} used the condition that the asymptotic behaviour of the $\Psi_0$ component of the Weyl tensor is $O(r^{-5})$. If a slower decay assumption for $\Psi_0$ is adopted, the logarithmic terms will appear in the asymptotic expansions. In this paper, we referred to a spacetime which can be asymptotically expanded in terms of $1/r$ as a smooth asymptotically flat spacetime.

In the celebrated work of the nonlinear stability of the Minkowski spacetime, Christodoulou and Klainerman proved that a generic asymptotically flat spacetime may admit a slower decay with $\Psi_0$ falling off as $O(r^{-7/2})$ \cite{Christo1993}. This scenario strongly suggests that the purely $1/r$ expansions are not adequately general. The limitations of the $1/r$ expansions  have been  noticed by many researchers and it seems natural to combine the powers of both $1/r$ and $\ln r$ in the asymptotic expansions \cite{Fock57,Bicak93,Ashtekar97,Chrusciel95,Kroon98,Kroon99}. Roughly speaking, a spacetime which can be expanded asymptotically in terms of $r^{-i}(\ln r)^j$ is called a polyhomogeneous spacetime, cf. \cite{Chrusciel95} for the precise definition. Chru\'{s}ciel, MacCallum and Singleton systematically analyzed the structure of polyhomogeneous spacetimes by a Bondi-Sachs type method \cite{Chrusciel95}. They showed that the assumption of polyhomogeneity is formally consistent with the Einstein equations. Later, Valiente Kroon managed to construct the logarithmic Newman-Penrose constants for polyhomogeneous spacetimes \cite{Kroon98,Kroon99}. Gasperin and Valiente Kroon investigated the relations between the polyhomogeneous expansions and the freely specifiable parts of time symmetric initial data \cite{Kroon17}. Fursaev {\it et al.} found that a spacetime generated by the cosmic strings admits asymptotically polyhomogeneous form \cite{Fursaev23}. Recently, Geiller, Laddha and Zwikel studied the symmetries of  the gravitational scattering in the absence of peeling within the Bondi-Sachs framework \cite{Geiller24}.

The notion of mass in general relativity is a quite subtle issue. Iyer and Wald established an effective framework to naturally introduce conserved quantities, including the mass  \cite{Iyer94}. This formalism works well in various diffeomorphism-covariant theories of gravity. Within the Iyer-Wald formalism, it was shown that the Arnowitt-Deser-Misner (ADM) mass is simply the Hamiltonian conjugate to the asymptotic time translation in spacetimes that are asymptotically flat at spatial infinity. To study the conserved quantities at null infinity, Wald {\it et al.} introduced a modified Hamiltonian and discussed the BMS charges in the conformal (unphysical) spacetime \cite{Wald2000, Grant2022}. In the current paper, we reexamine the Noether charge associated with the asymptotic time translation at null infinity in a direct way in the physical spacetime and no conformal compactification is required. It is shown that, for smooth asymptotically flat spacetimes, the Noether charge associated with the asymptotic time translation turns out to be the Bondi mass in the BS framework and the Newman-Unti (NU) mass respectively in the Newman-Penrose formalism. Moreover, the notion of mass at null infinity in polyhomogeneous spacetimes is discussed.

Gravitational wave memory is a significant prediction of general relativity which was firstly found by Zel'dovich and Polnarev \cite{Zel74}. This type of memory is produced by the change of quadrupole moments and is referred to as the linear memory. In the 1990s, Christodoulou and Frauendiener found the nonlinear memory of gravitational waves \cite{Christo91,Frau92}. Recently, for smooth asymptotically flat spacetimes, Nichols {\it et al.} proposed a BMS method to calculate the gravitational wave memory \cite{Nichols17,Khera21}. Cao {\it et al.} applied the BMS method to calculate the gravitational memory produced by binary black hole coalescence \cite{Cao21-1,Cao21-2}. One of the essential steps  in the BMS method is deriving the balance law \cite[Eq.(15)]{Cao21-1}, which relates the mass aspect to the asymptotic shear and thereby shows how gravitational radiation reduces the Bondi mass. The gravitational wave memory and the balance law for polyhomogeneous spacetimes are discussed in this paper. It is shown that the appearance of the logarithmic terms does not affect the balance law and it remains unchanged as the one that for smooth asymptotically flat spacetimes.

The organization of this paper is as follows. In Section \ref{s2}, we make use of the NP formalism to derive the asymptotic behavior of NP quantities, including the metric coefficients, the spin coefficients, and the tetrad components of the Weyl tensor for vacuum polyhomogeneous spacetimes. Section \ref{s3} focuses on the investigation of the asymptotic symmetries of polyhomogeneous spacetimes.
In Section \ref{s4},we review the Iyer-Wald formalism and reexamine the charge associated with asymptotic time translations at null infinity for smooth asymptotically flat spacetimes, utilizing a direct approach avoiding conformal compactifications.  In Section \ref{s5}, we address the problem of how to generalize the Bondi mass to polyhomogeneous spacetimes  via the Iyer-Wald formula. In Section \ref{s6}, the gravitational wave memory and the balance law of polyhomogeneous spacetimes are established. Conclusions are presented in Section \ref{s7}.

We adopt the geometric units $c=G=1$  throughout this paper. Following the convention established in the Newman-Penrose formalism \cite{NP62}, we choose to use the Landau-Lifshitz timelike convention for the signature of the spacetime metric, $(+,-,-,-)$. However, we shall make one exception to this convention in Section \ref{s4} to have the metric signature $(-,+,+,+)$ which Bondi {\it et al.} used. This is because it is more convenient for readers to revisit the relevant formulae for smooth asymptotically flat spacetimes in  \cite{Bondi62, Sachs62, CWYcmp2021}.

\section{Asymptotic behaviour of Newman-Penrose quantities}\label{s2}

We start with introducing a family of null hypersurfaces parameterised
by $u=\mbox{constant}$ with $u$ satisfying the Eikonal
equation
\begin{eqnarray}
g^{ab}\nabla_au\nabla_bu=0.\nonumber
\end{eqnarray}
Then we take $u$ as the $x^0$ coordinate and the first null vector will
be chosen as
\begin{eqnarray}
l_a=\nabla_a u.\nonumber
\end{eqnarray}
Since the $\{u=\mbox{constant}\}$ hypersurfaces are null, the $l_a$'s are tangent to a family of
curves lying within the hypersurfaces. These curves are null
geodesics and their tangent vectors $l^a$ satisfy
\begin{eqnarray}
l^a\nabla_al^b=0.\nonumber
\end{eqnarray}
Associated with these null geodesics lying in the $\{u=\mbox{constant}\}$hypersurfaces, an
affine parameter $r$ can be taken as the $x^1$ coordinate. The two
remaining coordinate $x^2=\theta, x^3=\varphi$ will label the geodesics.

We follow the notions in \cite{NP62}. In addition to the null vector $l^a$, another null
vector $n^a$ and the other two complex null vectors $m^a,\bar{m}^a$ can be written as
\begin{equation}
\begin{split}
l^a&=(\frac{\partial}{\partial r})^a,\\
n^a&=(\frac{\partial }{\partial u})^a+U(\frac{\partial}{\partial r})^a+X^{\theta}(\frac{\partial}{\partial\theta})^a
+X^{\varphi}(\frac{\partial}{\partial\varphi})^a,\\
m^a&=\omega(\frac{\partial}{\partial r})^a+\xi^{\theta}(\frac{\partial}{\partial\theta})^a+
\xi^{\varphi}(\frac{\partial}{\partial\varphi})^a,\\
\bar{m}^a&=\bar{\omega}(\frac{\partial}{\partial
r})^a+\bar{\xi}^{\theta}(\frac{\partial}{\partial\theta})^a
+\bar{\xi}^{\varphi}(\frac{\partial}{\partial\varphi})^a.
\end{split}
\ee
where $U,X^A (A=\theta,\varphi)$ are real quantities, and $\omega, \xi^A $ are complex quantities. All these quantites are functions of $u,r,\theta$ and $\varphi$.

The directional derivatives are defined as
\be
\begin{split}
D&=l^a\nabla_a=\frac{\partial}{\partial r},\\
\Delta&=n^a\nabla_a=\frac{\partial }{\partial u}+U\frac{\partial}{\partial r}+X^2\frac{\partial}{\partial\theta}
+X^3\frac{\partial}{\partial\varphi},\\
\hat{\delta}&=m^a\nabla_a=\omega\frac{\partial}{\partial r}+\xi^2\frac{\partial}{\partial\theta}+
\xi^3\frac{\partial}{\partial\varphi},\\
\bar{\hat{\delta}}&=\bar{m}^a\nabla_a=
\bar{\omega}\frac{\partial}{\partial
r}+\overline{\xi^2}\frac{\partial}{\partial\theta}
+\overline{\xi^3}\frac{\partial}{\partial\varphi}.
\end{split}
\ee
Here we use $\hat{\delta}$ to denote the directional derivative along $m^a$, and the notation $\delta$ is reserved to represent the variation in Sections \ref{s4} and \ref{s5}.

For later convenience of calculation, we adopt the following
gauge on the NP tetrad
\begin{eqnarray}
\kappa=\epsilon=\pi=0,\ \rho=\bar{\rho},\ \tau=\bar{\alpha}+\beta,\label{G1}
\end{eqnarray}
where $\kappa,\epsilon,\pi,\rho,\tau,\alpha$ and $\beta$ are spin coefficients \cite{NP62}.
This gauge condition follows from the fact that $l_a=\nabla_a u$ is a
gradient field and it implies that the tetrad vectors $m^a$ and $n^a$ are chosen to be parallel-propagated along $l^a$. Note that the gauge choice \eqref{G1}  we adopt here is slightly different with the one used in \cite[Eqs.(13) and (14)]{Kroon99}.

Within the Newman-Penrose formalism, the 38 NP equations (including the metric equations, the spin coefficient equations and the Bianchi identities) can be solved order-by-order. For this purpose, we need $\Psi_0$ on an initial null hypersurface $\mathcal{N}_0$ \cite{Kroon98}.

The most general polyhomogeneous form for the physically reasonable  $\Psi_0$ \cite{Kroon99} is
\be
\Psi_0=\sum_{k=3}\Psi_0^kr^{-k},
\ee
where
\be
\Psi_0^k=\sum_{j=0}^{N_k}\Psi_0^{k,j}z^{j},
\ee
with
$ z=\ln r$.

When solving the NP equations, one usually needs to calculate both the derivatives and the products of the polyhomogeneous functions \cite{Kroon98}. Let
$h$ and $g$ be two polyhomogeneous functions as
\be
\begin{split}
h&=\sum_{i=1}h_i(z)r^{-i},\\
g&=\sum_{i=1}g_i(z)r^{-i},
\end{split}
\ee
where the coefficients $h_i(z), g_i(z)$ are polynomials in $z=\ln r$. Then we have
\be
\begin{split}
hg&=\sum_{i=2}\sum_{k=1}^{i-1}h_k(z)g_{i-k}(z)r^{-i},\\
\frac{\partial}{\partial r}h&=\sum_{i=2}\bigg[
h'_{i-1}(z)-(i-1)h_{i-1}(z)\bigg]r^{-i},
\end{split}
\ee
where $'$ denotes the differentiation with respect to $z$.

For polyhomogenous spacetimes, all the NP quantities can be expanded in terms of $r^{-i}(\ln r)^j$. For instance,
\be
\begin{split}
\rho=\sum_{i=1}\rho_ir^{-i}=\sum_{i=1}(\sum_{k=0}^{}\rho_{i,k}z^k)r^{-i},\\
\sigma=\sum_{i=1}\sigma_ir^{-i}=\sum_{i=1}(\sum_{k=0}^{}\sigma_{i,k}z^k)r^{-i},
\end{split}
\ee
where $\rho_{i,k}$ and $\sigma_{i,k}$ are functions of $u,\theta$ and $\varphi$. We will denote by $\#$ the degree of a polynomial in $z$  and use the convention that the zero polynomial has degree $-\infty$.

In the following, we will solve the NP equations order-by-order. The procedure is similar to the one done in
\cite{NP62,Kroon99}. We only give the details for the solving process of the first two equations in the NP hierarchy which are
\be\label{NP1,2}
\begin{split}
 D\rho&=\rho^2+\sigma\bar{\sigma},\\
 D\sigma&=2\rho\sigma+\Psi_0.
\end{split}
\ee
From the above two equations we can get the following recurrence relations
\be\label{np1,2}
\begin{split}
\rho'_{i-1}-(i-1)\rho_{i-1}&=\sum_{k=1}^{i-1}[
\rho_k\rho_{i-k}+\sigma_k\bar\sigma_{i-k}],\\
\sigma'_{i-1}-(i-1)\sigma_{i-1}&=2\sum_{k=1}^{i-1}\rho_k\sigma_{i-k}+\Psi_0^i.
\end{split}
\ee

For $i=2$, the recurrence relations are reduced to
\be
\begin{split}\label{rho1}
\rho_1'-\rho_1&=\rho_1^2+\sigma_1\bar{\sigma}_1,\\
\sigma_1'-\sigma_1&=2\rho_1\sigma_1,
\end{split}
\ee
since $\Psi_0^2=0.$
This leads to the consequence that $\rho_1$ and $\sigma_1$ must be polynomials of degree zero, which in turn yields
\be
\begin{split}
-\rho_1&=\rho_1^2+\sigma_1\bar{\sigma}_1,\\
-\sigma_1&=2\rho_1\sigma_1.
\end{split}
\ee
The above equations imply that
\be
\rho_1=-1,\sigma_1=0.
\ee

For $i=3$, (\ref{np1,2}) gives
\be
\begin{split}
\rho_2'&=0,\\
\sigma_2'&=\Psi_0^3.
\end{split}
\ee
Hence $\rho_2$ is a polynomial of degree zero in $z$. By the affine freedom in the definition of the radial coordinate, $\rho_2$ can be set to zero if necessary. Noting that
\be
\sigma_2'=\sum_j(j+1)\sigma_{2,j+1}z^j,\ \ \Psi_0^3=\sum_j\Psi_0^{3,j}z^j,
\ee
we have
\be
(j+1)\sigma_{2,j+1}=\Psi_0^{3,j},
\ee
where $j=0,1,\cdots, N_3$, whence
 $$^{\#}\sigma_2=N_3+1$$
and
\be\label{2.18}
\sigma_{2,j+1}=\frac{\Psi_0^{3,j}}{j+1},\ \ (j=0,1,\cdots, N_3).
\ee
All the coefficients of $\sigma_2$ can be determined by $\Psi_0^3$ on $\mathcal{N}_0$ except for $\sigma_{2,0}$.

For $i=4$, (\ref{np1,2}) gives
\be\label{rho3}
\begin{split}
\rho_3'-\rho_3=\sigma_2\bar{\sigma}_2,\\
\sigma_3'-\sigma_3=\Psi_0^4
\end{split}
\ee
and
\be
^{\#}\rho_3=2N_3+2,\ \ \ ^{\#}\sigma_3=N_4.
\ee
The first equation in (\ref{rho3}) yields
\be
\begin{split}
(j+1)\rho_{3,j+1}-\rho_{3,j}&=\sum_{k=0}^j\sigma_{2,k}\bar{\sigma}_{2,j-k},\ \
 \ (j=0,1,\cdots, 2N_3+1)\\
-\rho_{3,2N_3+2}&=\sigma_{2,N_3+1}\bar{\sigma}_{2,N_3+1}.
\end{split}
\ee
This implies that $\rho_3(z)$ is determined by $\sigma_2(z)$ and further it is completely determined by the initial data $\Psi_0^3(z)$ and $\sigma_{2,0}$.

The second equation in (\ref{rho3}) gives
\be
\begin{split}
(j+1)\sigma_{3,j+1}-\sigma_{3,j}&=\Psi_0^{4,j},\ \ (j=0,1,\cdots,N_4-1)\\
N_4\sigma_{3,N_4}&=\Psi_0^{4,N_4}.
\end{split}
\ee
It follows  that $\sigma_3(z)$ is completely determined by the initial data $\Psi_0^4(z).$

For $i=5$, (\ref{np1,2}) gives
\be\label{rho4}
\begin{split}
\rho_4'-2\rho_4&=\sigma_2\bar{\sigma}_3+\sigma_3\bar{\sigma}_2,\\
\sigma_4'-2\sigma_4&=2\rho_3\sigma_2+\Psi_0^5.
\end{split}
\ee
From these two equations, one arrives at
\be
^{\#}\rho_4=N_3+N_4+1,\ \ \ ^{\#}\sigma_4=\max\{3N_3+3,N_5\}.
\ee
The first equation in (\ref{rho4}) gives
\be
\begin{split}
(j+1)\rho_{4,j+1}-2\rho_{4,j}&=
\sum_{k=0}^{j}(\sigma_{2,k}\bar{\sigma}_{3,j-k}
+\bar\sigma_{2,k}\sigma_{3,j-k}),\ \ (j=0,1,\cdots,N_3+N_4)\\
-2\rho_{4,N_3+N_4+1}&=\sigma_{2,N_3+1}\bar{\sigma}_{3,N_4}+
\bar{\sigma}_{2,N_3+1}\sigma_{3,N_4}.
\end{split}
\ee
This implies that $\rho_4(z)$ is completely determined by $\sigma_2(z)$ and
$\sigma_3(z)$, which in turn is determined essentially by the initial data $\Psi_0^3(z)$, $\sigma_{2,0}$ and $\Psi_0^4(z)$.

The second equation in (\ref{rho4}) gives
\be
\begin{split}
(j+1)\sigma_{4,j+1}&-2\sigma_{4,j}
=2\sum_{k=0}^j(\rho_{3,k}\sigma_{2,j-k})+\Psi_0^{5,j},\\
& \  \ (j=0,1,\cdots,\max\{3N_3+3,N_5\}-1)
\end{split}
\ee
when $N_5=3N_3+3,$
\be
N_5\sigma_{4,N_5}=2\rho_{3,2N_3+2}\sigma_{2,N_3+1}+\Psi_0^{5,N_5};
\ee
when $N_5>3N_3+3,$
\be
N_5\sigma_{4,N_5}=\Psi_0^{5,N_5};
\ee
when $N_5<3N_3+3,$
\be
(3N_3+3)\sigma_{4,3N_3+3}=2\rho_{3,2N_3+2}\sigma_{2,N_3+1}.
\ee
Therefore $\sigma_4(z)$ is determined by $\rho_3(z),\sigma_2(z)$ and
$\Psi_0^5(z),$ which in turn is determined essentially by the initial data $\sigma_{2,0}$, $\Psi_0^3(z)$ and $\Psi_0^5(z)$.

The above process can be continued up to any desired order in $1/r$. For the remaining NP equations, similar analysis can be carried out. We summarize the final results as follows.

The asymptotic behaviour of the spin coefficients are
\be
\begin{split}
 \sigma=&\frac{\sigma_2(z)}{r^2}+\frac{\sigma_3(z)}{r^3}
+\frac{\sigma_4(z)}{r^4}+\cdots,\\
&  ^{\#} \sigma_2=N_3+1,\ ^{\#}\sigma_3=N_4, \ ^{\#}\sigma_4=\max\{3N_3+3,N_5\},\cdots,\\
&\sigma_2=\sigma_{2,0}(u,\theta,\varphi)+\sum_{i=1}^{N_3+1}
\sigma_{2,i}(\theta,\varphi)z^i,\\
&\sigma_{2,j}=\frac{1}{j}\Psi_{0}^{3,j},\ (j=1,2,\cdots,N_3+1).
\end{split}
\ee
\be
\begin{split}
 \rho=&-\frac{1}{r}+\frac{\rho_3(z)}{r^3}+\frac{\rho_4(z)}{r^4}+\cdots,\\
&^{\#}\rho_3=2N_3+2, \ ^{\#}\rho_4=N_3+N_4+1,\cdots,\\
&\rho_3=\sum_{j=0}^{2N_3+2}\rho_{3,j}z^j,\\
&\rho_{3,2N_3+2}=-\sigma_{2,N_3+1}\bar\sigma_{2,N_3+1},\\
&\rho_{3,j}=(j+1)\rho_{3,j+1}-\sum_{k=0}^j\sigma_{2,k}\bar\sigma_{2,j-k},\ (j=2N_3+1,\cdots,1,0).
\end{split}
\ee
\be
\begin{split}
 \alpha=&\frac{\alpha_1}{r}+
\frac{\alpha_2(z)}{r^2}+\cdots,\ \ \ ^{\#}\alpha_1=0,\
^{\#}\alpha_2=N_3+1,\cdots,\\
&\alpha_1=\alpha_{1,0}=-\frac{\cot\theta}{2\sqrt{2}},\\
&\alpha_2=\sum_{j=0}^{N_3+1}\alpha_{2,j}z^j,\\
&\alpha_{2,N_3+1}=-\beta_{1,0}\bar\sigma_{2,N_3+1},\ \ \beta_{1,0}=\frac{\cot\theta}{2\sqrt{2}},\\
&\alpha_{2,j}=(j+1)\alpha_{2,j+1}-\beta_{1,0}\bar\sigma_{2,j},\ (j=N_3,\cdots,1,0).\ \ \ \ \ \ \ \ \ \ \ \ \
\end{split}
\ee
\be
\begin{split}
 \beta=&\frac{\beta_1}{r}+
\frac{\beta_2}{r^2}+\cdots,\ \ \ ^{\#}\beta_1=0,\
^{\#}\beta_2=N_3+1,\cdots,\\
&\beta_1=\beta_{1,0}=\frac{\cot\theta}{2\sqrt{2}},\\
&\beta_2=\sum_{j=0}^{N_3+1}\beta_{2,j}z^j,\\
&\beta_{2,N_3+1}=-\alpha_{1,0}\sigma_{2,N_3+1},\\
&\beta_{2,j}=(j+1)\beta_{2,j+1}-\alpha_{1,0}\sigma_{2,j}-\Psi_1^{3,j},\ (j=N_3,\cdots,1,0).\ \ \
\end{split}
\ee
\be
\begin{split}
 \tau=&\frac{\tau_2}{r^2}+\frac{\tau_3}{r^3}+\cdots,\ \ \ ^{\#}\tau_2=N_3,\cdots,\\
&\tau_2=\sum_{j=0}^{N_3}\tau_{2,j}z^j,\\
&\tau_{2,j}=\bar\alpha_{2,j}+\beta_{2,j},\ (j=0,1,\cdots,N_3).\ \ \ \ \ \ \ \ \ \ \ \ \ \ \ \ \ \ \ \ \ \ \ \ \ \ \ \
\end{split}
\ee
\be
\begin{split}
 \lambda=&\frac{\lambda_1}{r}+\frac{\lambda_2}{r^2}+\cdots,\ \ \ ^{\#}\lambda_1=0,\ ^{\#}\lambda_2
=N_3+1,\cdots,\\
&\lambda_1=\dot{\bar\sigma}_{2,0},\\
&\lambda_2=\sum_{j=0}^{N_3+1}\lambda_{2,j}z^j,\\
&\lambda_{2,N_3+1}=-\mu_{1,0}\bar\sigma_{2,N_3+1}=\frac{1}{2}\bar\sigma_{2,N_3+1},\\
&\lambda_{2,j}=(j+1)\lambda_{2,j+1}+\frac{1}{2}\bar{\sigma}_{2,j},\ (j=N_3,\cdots,1,0).\ \ \ \ \ \ \ \ \ \ \ \ \
\end{split}
\ee
\be
\begin{split}
 \mu=&\frac{\mu_1}{r}+\frac{\mu_2}{r^2}+\cdots,\ \ \ ^{\#}\mu_1=0,\ ^{\#}\mu_2=N_3+1,\cdots,\\
&\mu_1=-\frac{1}{2},\\
&\mu_2=\sum_{j=0}^{N_3+1}\mu_{2,j}z^{j},\\
&\mu_{2,N_3+1}=-(\lambda_{1,0}\sigma_{2,N_3+1}+\Psi_2^{3,N_3+1}),\\
&\mu_{2,j}=(j+1)\mu_{2,j+1}-\sigma_{2,j}\lambda_{1,0}-\Psi_{2}^{3,j},\ (j=N_3,\cdots,1,0).
\end{split}
\ee
\be\label{2.40}
\begin{split}
 \gamma=&\frac{\gamma_2}{r^2}+\frac{\gamma_3}{r^3}+\cdots,\ \ \ \ ^{\#}\gamma_2=N_3+1,\cdots,\\
&\gamma_2=\sum_{j=0}^{N_3+1}\gamma_{2,j}z^j,\\
&\gamma_{2,N_3+1}=-\frac{1}{2}\Psi_2^{3,N_3+1},\\
&\gamma_{2,j}=\frac{1}{2}\bigg[(j+1)\gamma_{2,j+1}-\alpha_{1,0}\tau_{2,j}-
\beta_{1,0}\bar\tau_{2,j}-\Psi_2^{3,j}\bigg],(j=N_3,\cdots,1,0).
\end{split}
\ee
\be
\begin{split}
 \nu=&\frac{\nu_1}{r}+\frac{\nu_2}{r^2}+\cdots,\ \ \ ^{\#}\nu_1=0,\ ^{\#}\nu_2=N_3+1,\cdots,\\
&\nu_1=-\Psi_3^{2,0}=\eth\dot{\bar\sigma}_{2,0},\\
&\nu_2=\sum_{j=0}^{N_3+1}\nu_{2,j}z^j,\\
&\nu_{2,N_3+1}=-\frac{1}{2}\Psi_3^{3,N_3+1},\\
&\nu_{2,j}=\frac{1}{2}\bigg[(j+1)\nu_{2,j+1}+\frac{1}{2}\bar\tau_{2,j}-\lambda_{1,0}\tau_{2,j}-\Psi_3^{3,j}\bigg],\ (j=N_3,\cdots,1,0).
\end{split}
\ee

The asymptotics of the metric coefficients are
\be\label{2.42}
\begin{split}
 U=&U_0+\frac{U_1}{r}+\frac{U_2}{r^2}+\cdots,\ \ \ ^{\#}U_0=0,\ ^{\#}U_1=N_3+1,\cdots,\\
&U_0=-\frac{1}{2},\\
&U_1=\sum_{j=0}^{N_3+1}U_{1,j}z^j,\\
&U_{1,N_3+1}=\gamma_{2,N_3+1}+\bar\gamma_{2,N_3+1},\\
&U_{1,j}=(j+1)U_{1,j+1}+\gamma_{2,j}+\bar\gamma_{2,j},\ (j=N_3,\cdots,1,0).
\end{split}
\ee
\be
\begin{split}
 X^A=&\frac{X_2^A}{r^2}+\cdots,\ \ \ \ ^{\#}X_2^A=N_3,\cdots\\
&X_2^A=\sum_{j=0}^{N_3}X_{2,j}^Az^j,\\
&X_{2,N_3}^A=-\frac{1}{2}\bigg[\bar\tau_{2,N_3}\xi^A_{1,0}+\tau_{2,N_3}\bar\xi^A_{1,0}\bigg],\\
&X_{2,j}^A=\frac{1}{2}\bigg[(j+1)X^A_{2,j+1}-\bar{\tau}_{2,j}\xi^A_{1,0}-\tau_{2,j}\bar\xi^A_{1,0}\bigg],\ (j=N_3-1,\cdots,1,0).
\end{split}
\ee
\be
\begin{split}
 \omega=&\frac{\omega_1(z)}{r}+\frac{\omega_2}{r^2}+\cdots,\ \ \ ^{\#}\omega_1=N_3+1,\cdots,\ \ \ \ \ \ \ \ \ \ \ \ \ \ \ \ \ \ \ \ \ \ \ \  \ \ \ \\
&\omega_1=\sum_{j=0}^{N_3+1}\omega_{1,j}z^j,\\
&\omega_{1,j+1}=-\frac{1}{j+1}(\bar\alpha_{2,j}+\beta_{2,j}),\ (j=0,1,\cdots,N_3),\\
&\omega_{1,0}=\bar{\eth}\sigma_{2,0}-\bar{\eth}\Psi_0^{3,0}+\bar{\eth}\sigma_{2,1}+2\omega_{1,2},\\
&\dot\omega_{1,0}=\bar{\eth}\dot\sigma_{2,0},\ \dot\omega_{1,2}=\frac{1}{2}\bar{\eth}\dot\Psi_0^{3,0}.
\end{split}
\ee

\be\label{xiA}
\begin{split}
 \xi^{A}=&\frac{\xi^A_{1}}{r}+\frac{\xi_2^{A}(z)}{r^2}+\cdots,\ \ ^{\#}\xi^A_1=0, \ ^{\#}\xi^{A}_2=N_3+1,\cdots,\ \ \ \ \ \   \ \ \ \ \ \ \ \\
&\xi_1^{\theta}=\xi_{1,0}^{\theta}=\frac{1}{\sqrt{2}},\\
&\xi_1^{\varphi}=\xi_{1,0}^{\varphi}=\frac{i}{\sqrt{2}\sin\theta},\\
&\xi_2^{A}=\sum_{j=0}^{N_3+1}\xi_{2,j}^{A}z^j,\\
&\xi_{2,N_3+1}^{A}=-\sigma_{2,N_3+1}\bar\xi_{1,0}^{A},\\
&\xi_{2,j}^{A}=(j+1)\xi_{2,j+1}^{A}-\sigma_{2,j}\bar\xi_{1,0}^{A},\ (j=N_3,\cdots,1,0).
\end{split}
\ee

The asymptotics of the tetrad components of the Weyl tensor are
\be\label{2.46}
\begin{split}
 \Psi_0=&\frac{\Psi_0^3(z)}{r^3}+\frac{\Psi_0^4(z)}{r^4}
+\cdots,\ \ \ \ ^{\#}\Psi_0^3=N_3,\ ^{\#}\Psi_0^4=N_4,\cdots,\\
&\frac{\partial\Psi_0^3}{\partial u}=0,\\
&\frac{\partial\Psi_0^4}{\partial u}=\frac{1}{2}(\Psi_0^3)'-\Psi_0^3+\eth\Psi_1^3.
\end{split}
\ee
\be\label{2.47}
\begin{split}
 \Psi_1=&\frac{\Psi_1^3(z)}{r^3}+
\frac{\Psi_1^4(z)}{r^4}+\cdots,\\
&  ^{\#}\Psi_1^3=N_3,\ ^{\#}\Psi_1^4=\max\{N_4+1,2N_3+2\}\cdots,\\
&\Psi_1^3=\sum_{j=0}^{N_3}\Psi_1^{3,j}z^j,\\
&\Psi_1^{3,N_3}=\bar{\eth}\Psi_0^{3,N_3},\\
&\Psi_1^{3,j}=-(j+1)\Psi_1^{3,j+1}+\bar{\eth}\Psi_0^{3,j},\ (j=N_3-1,\cdots,1,0).\ \
\end{split}
\ee
\be\label{2.48}
\begin{split}
 \Psi_2=&\frac{\Psi_2^3}{r^3}+\cdots,\ \ \ \ ^{\#}\Psi_2^3=N_3+1,\dots\\
&\Psi_2^3=\sum_{j=0}^{N_3+1}\Psi_2^{3,j}z^j,\\
&\Psi_2^{3,j}=\frac{1}{j}\bigg[\bar{\eth}\Psi_1^{3,j-1}-\lambda_{1,0}\Psi_0^{3,j-1}\bigg],(j=1,2,\cdots,N_3+1),\\
&\dot{\Psi}_2^3=\eth\Psi_3^2+\sigma_2\Psi_4^1,\\
&\frac{\partial\Psi_2^{3,0}}{\partial u}=-\eth^2\dot{\bar{\sigma}}_{2,0}-\sigma_{2,0}\ddot{\bar\sigma}_{2,0}.
\end{split}
\ee
\be
\begin{split}
 \Psi_3=&\frac{\Psi_3^2}{r^2}+\frac{\Psi_3^3}{r^3}+\cdots,\ \ \ ^{\#}\Psi_3^2=0,\
^{\#}\Psi_3^3=N_3+1,\cdots,\\
&\Psi_3^2=\Psi_3^{2,0}=-\eth\dot{\bar\sigma}_{2,0},\\
&\Psi_3^3=\sum_{j=0}^{N_3+1}\Psi_3^{3,j}z^j,\\
&\Psi_3^{3,N_3+1}=-\bar{\eth}\Psi_2^{3,N_3+1},\\
&\Psi_3^{3,j}=(j+1)\Psi_3^{3,j+1}-\bar{\eth}\Psi_2^{3,j}+2\lambda_{1,0}\Psi_1^{3,j},\ (j=N_3,\cdots,1,0).
\end{split}
\ee
\be
\begin{split}
 \Psi_4=&\frac{\Psi_4^1}{r}+\frac{\Psi_4^2}{r^2}+\cdots,\ \ \
^{\#}\Psi_4^1=0,\ ^{\#}\Psi_4^2=0,\cdots,\\
&\Psi_4^1=-\ddot{\bar\sigma}_{2,0},\\
&\Psi_4^2=-\bar{\eth}\Psi_3^{2,0}=\bar{\eth}\eth\dot{\bar\sigma}_{2,0}.\ \ \ \ \ \ \ \ \ \ \ \ \ \ \ \ \ \ \ \ \ \ \ \ \ \ \ \ \ \ \ \ \ \ \ \ \ \ \
\end{split}
\ee
Here $\eth$ and $\bar{\eth}$ are the spin-weighted operators \cite{Penrose88}.

For later usage, we  list the expression of $\xi_2^{\theta}$ and $\xi_2^{\varphi}$ here:
\be\label{xithetaphi}
\begin{split}
\xi^{\theta}_{2,N_3+1}=&-\frac{1}{\sqrt{2}}\sigma_{2,N_3+1},\\
(j+1)\xi^{\theta}_{2,j+1}-\xi^{\theta}_{2,j}=&\frac{1}{\sqrt{2}}\sigma_{2,j},
\ \ (N_3=0,1,\cdots, N_3),\\
\xi^{\varphi}_2=&-\frac{i}{\sin\theta}\xi^{\theta}_2.
\end{split}
\ee
The above formulae can be obtained from \eqref{xiA}.

\section{Asymptotic symmetries of polyhomogeneous spacetimes}\label{s3}

In this section, we study the asymptotic symmetries of the polyhomogenous spactimes.
From the relation
\be
g^{ab}=l^an^b+n^al^b-m^a\bar m^b-\bar m^am^b,
\ee
one can get the contravariant metric components $g^{\mu\nu}$ within the NU coordinates $\{u, r,\theta,\varphi\}$ as
\be
g^{\mu\nu}=\left(
             \begin{array}{cccc}
               0 & 1 & 0 & 0 \\
               1 & 2U-2\omega\bar\omega & X^{\theta}-\omega\bar\xi^{\theta}
                -\bar\omega\xi^{\theta}& X^{\varphi}-\omega\bar\xi^{\varphi}-\bar\omega\xi^{\varphi} \\
               0 & X^{\theta}-\omega\bar\xi^{\theta}
                -\bar\omega\xi^{\theta} & -2\xi^{\theta}\bar\xi^{\theta} & -\xi^{\theta}\bar\xi^{\varphi}-\bar\xi^{\theta}\xi^{\varphi} \\
               0 & X^{\varphi}-\omega\bar\xi^{\varphi}-\bar\omega\xi^{\varphi} & -\xi^{\theta}\bar\xi^{\varphi}-\bar\xi^{\theta}\xi^{\varphi} & -2\xi^{\varphi}\bar\xi^{\varphi} \\
             \end{array}
           \right).
\ee
More explicitly, by using the asymptotics obtained in last section, we have
\be
\begin{split}
g^{11}=&-1+\frac{2U_1}{r}+\frac{2U_2-2\omega_1\bar\omega_1}{r^2}+\cdots,\\
g^{12}=&\bigg(X_2^{\theta}-\frac{\omega_1}{\sqrt{2}}
-\frac{\bar\omega_1}{\sqrt{2}}\bigg)\frac{1}{r^2}+\cdots,\\
g^{13}=&\bigg(X_2^{\varphi}+\frac{i\omega_1}{\sqrt{2}\sin\theta}
-\frac{i\bar\omega_1}{\sqrt{2}\sin\theta}\bigg)\frac{1}{r^2}+\cdots,\\
g^{22}=&-\frac{1}{r^2}-\sqrt{2}\big(\xi^{\theta}_2+\bar\xi^{\theta}_2
\big)\frac{1}{r^3}+\cdots,\\
g^{23}=&\bigg(\frac{i(\xi^{\theta}_2-
\bar\xi^{\theta}_2)}{\sqrt{2}\sin\theta}
-\frac{\xi^{\varphi}_2+\bar\xi^{\varphi}}{\sqrt{2}}
\bigg)\frac{1}{r^3}+\cdots,\\
g^{33}=&-\frac{1}{r^2\sin^2\theta}
+\frac{i\sqrt{2}}{\sin\theta}\bigg(\xi^{\varphi}_2-\bar\xi^{\varphi}_2
\bigg)\frac{1}{r^3}+\cdots.
\end{split}
\ee

Furthermore, straightforward calculation shows that the covariant metric components in the NU coordinate system are
\be\label{3.4}
\begin{split}
g_{00}&=1-\frac{2U_1}{r}+\cdots,\\
g_{01}&=1,\\
g_{02}&=\bigg(X^{\theta}_2-\frac{\omega_1+
\bar\omega_1}{\sqrt{2}}\bigg)+\cdots,\\
g_{03}&=\bigg(X^{\varphi}_2+\frac{i(\omega_1-
\bar\omega_1)}{\sqrt{2}\sin\theta}\bigg)\sin^2\theta+\cdots,\\
\end{split}
\ee
and
\be\label{3.5}
\begin{split}
g_{11}&=0,\\
g_{12}&=0,\\
g_{13}&=0,\\
g_{22}&=-r^2+\sqrt{2}(\xi^{\theta}_2+\bar\xi^{\theta}_2)r+\cdots,\\
g_{23}&=\frac{\sin\theta}{\sqrt{2}}\bigg(
i\bar\xi^{\theta}_2-i\xi^{\theta}_2+\sin\theta(\xi^{\varphi}_2
+\bar\xi^{\varphi}_2)\bigg)r+\cdots,\\
g_{33}&=-r^2\sin^2\theta-i\sqrt{2}
\sin^3\theta(\xi^{\varphi}_2-\bar\xi^{\varphi}_2)r+\cdots.\\
\end{split}
\ee

Let $\xi^{\mu}$ be a vector field that generating an infinitesimal transformation preserving the metric forms of (\ref{3.4}) and (\ref{3.5}), i.e.,
\be
\begin{split}
\mathcal{L}_{\xi}g_{uu}&=O\big(r^{-1}(\ln r)^{N_3+1}\big),\\
\mathcal{L}_{\xi}g_{ur}&=\mathcal{L}_{\xi}g_{rr}
=\mathcal{L}_{\xi}g_{rA}=0,\\
\mathcal{L}_{\xi}g_{uA}&=O\big((\ln r)^{N_3+1}\big),\\
\mathcal{L}_{\xi}g_{AB}&=O\big(r(\ln r)^{N_3+1}\big).
\end{split}
\ee
Solving the above asymptotic Killing equations, we find
\be\label{3.7}
\begin{split}
\xi^u=&f(\theta,\varphi)+\frac{1}{2}(D_AY^A)u+\cdots,\\
\xi^r=&-\frac{1}{2}(D_AY^A)\ r+\cdots,\\
\xi^{\theta}=&Y^{\theta}+\cdots,\\
\xi^{\varphi}=&Y^{\varphi}+\cdots,
\end{split}
\ee
where $f(\theta,\varphi)$ is an arbitrary smooth function on the 2-sphere, $D_A$ stands for the covariant derivative of the standard unit 2-sphere, and $Y^A$
depends only on $\theta$ and $\varphi$ satisfying
\be\label{LAB}
\mathcal{L}_Yq_{AB}=\frac{1}{2}(D_CY^C)q_{AB}
\ee
with $q_{AB}$ the round metric on the unit  2-sphere.  Equation (\ref{LAB}) implies that $Y^A$ is actually a conformal Killing vector field on the round sphere. Moreover, if $\xi_1^a$ and $\xi^a_2$ are both asymptotic infinitesimal symmetries, i.e., the components of $\xi_1^a$ and $\xi^a_2$ admit the form (\ref{3.7}), direct calculation shows that $[\xi_1,\xi_2]^a$ is also an asymptotic infinitesimal symmetry. The above results are exactly the same as those obtained in the smooth asymptotically flat spacetime.

\section{Bondi mass of smooth asympotically flat spacetimes}\label{s4}

In this section, we reexamine the  charge associated with the asymptotic time translation at null infinity of smooth asymptotically flat spacetimes via the Iyer-Wald formula within the Bondi-Sachs coordinate system.

For the theory of general relativity, the Lagrange 4-form reads \cite{Iyer94}
\be
L_{abcd}=\frac{R}{16\pi}\epsilon_{abcd},
\ee
which yields a symplectic potential 3-form
\be\label{potential3form}
\Theta_{abc}=\epsilon_{dabc}\frac{1}{16\pi}g^{de}g^{fh}
(\nabla_f\delta g_{eh}-\nabla_e\delta g_{fh}).
\ee

For an asymptotic Killing vector field  $\xi^a$, one  has the associated Noether current 3-form
\be
J_{abc}=\frac{1}{8\pi}\epsilon_{dabc}\nabla_e(\nabla^{[e}\xi^{d]})
\ee
and the associated Noether charge 2-form
\be
Q_{ab}=-\frac{1}{16\pi}\epsilon_{abcd}\nabla^c\xi^d.
\ee

When $\xi^a$ is an asymptotic time translation, if there exists a 3-form $B_{abc}$ such that
\be
\delta\int_{\infty}\xi\cdot B=\int_{\infty}\xi\cdot\Theta,
\ee
Iyer and Wald defined the canonical energy as \cite[Eq.(83)] {Iyer94}

\be\label{waldmass-new}
\mathcal{E}\equiv\int_{\infty}Q-\xi\cdot B.
\ee
We make use of $\int_{\infty}$ to denote the integral over a 2-sphere at null infinity $\lim\limits_{r\rightarrow\infty}\int_{S_r}$ in the rest of this paper.

Within the Bondi-Sachs coordinate system $\{u,r,x^A\} (A=2,3)$, the line element of an asymptotically flat spacetime is given by
\cite[ Eq.(2.1)]{CWYcmp2021}
\begin{eqnarray}\label{ds-new-1}
\d s^2=-UV\d u^2-2U\d u\d r+r^2h_{AB}(\d x^A+W^A\d u)(\d x^B+W^B\d u),
\end{eqnarray}
where \cite[ Page 6]{CWYcmp2021}
\begin{equation}\label{ds-new-2}
\begin{split}
U&=1-\frac{1}{16r^2}C_{AB}C^{AB}+\cdots,\\
V&=1-\frac{2m}{r}+\frac{1}{r^2}\bigg(\frac{1}{3}D^AN_{A}+\frac{1}{4}(D^AC_{AB})(D_EC^{BE})\\
&+\frac{1}{16}C_{EF}C^{EF}\bigg)+\cdots, \\
W^A&= \frac{1}{2r^2}D_BC^{AB}+\frac{1}{r^3}\bigg(
\frac{2}{3}N^{A}_{}
-\frac{1}{16}D^{A}(C_{EF}C^{EF})\\
&
-\frac{1}{2}C^{AB}D^EC_{BE}\bigg)+\cdots, \\
h_{AB}&=q_{AB}+\frac{C_{AB}}{r}+\frac{q_{AB}}{4r^2}C_{EF}C^{EF}+\cdots.
\end{split}
\end{equation}
Here, the indices are contracted, raised, and lowered with respect to
 the round metric on the unit sphere $q_{AB}$, $m = m(u, x_A
)$ is the mass aspect, $N_A = N_A(u,
x_A
)$ is the angular aspect, and $C_{AB} = C_{AB}(u, x_A
)$ is the shear
tensor. The news tensor $N_{AB}$ is defined
as
\be
N_{AB}=\frac{\partial C_{AB}}{\partial u}.
\ee

It follows from Eqs.(\ref{ds-new-1}) and (\ref{ds-new-2}) that
\begin{equation}
\begin{split}
\d s^2&=\bigg(-1+\frac{2m}{r}\bigg)\d u^2-2\bigg(1-\frac{1}{16r^2}C_{AB}C^{AB}\bigg)\d u \d r\\
&+\bigg[D^E C_{AE}+\frac{1}{r}\bigg(\frac{4}{3}N_{A}-\frac{1}{8}D_A(C_{EF}C^{EF})\bigg)\bigg]\d u\d x^A\\
&+r^2\bigg(q_{AB}+\frac{C_{AB}}{r}+\frac{q_{AB}}{4r^2}C_{EF}C^{EF}\bigg)\d x^A\d x^B\\
&+(\mbox{subleading terms in}\  \frac{1}{r}).
\end{split}
\end{equation}

Furthermore, the
evolution equations of $m$ and $N_{A}$ are given by
\cite[Eqs.(3.1) and (3.4)]{CWYcmp2021}
\begin{equation}
\begin{split}
\partial_um&=-\frac{1}{8}N_{AB}N^{AB}+\frac{1}{4}D^AD^BN_{AB},\\
\partial_uN_{A}&=D_Am-\frac{1}{4}D^BD_BD^EC_{EA}+\frac{1}{4}D^BD_AD^EC_{EB}\\
&+\frac{1}{4}D_A(C_{BE}N^{BE})-\frac{1}{4}D_B(C^{BD}N_{DA})+\frac{1}{2}C_{AB}D_EN^{EB}.
\end{split}
\end{equation}

The variation of metric components has the following behaviour
\begin{eqnarray}\label{C2}
&& \delta g_{uu}\sim O(\frac{1}{r}),\\
&& \delta g_{ru}\sim O(\frac{1}{r^2}),\\
&& \delta g_{rr},\delta g_{r\theta},\delta g_{r\varphi}=0,\\
&& \delta g_{u\theta},\delta g_{u\varphi}\sim O(1),\\
&& \delta g_{\theta\theta},\delta g_{\theta\varphi},\delta g_{\varphi\varphi}\sim O(r).
\end{eqnarray}

Let $\xi^a$ be the asymptotic time translation
\be
\xi^a=(\frac{\partial}{\partial u})^a.
\ee
The 2-sphere at null infinity is referred as to  the limit as $r\rightarrow\infty$ of the coordinate spheres $\{r=constant,\ u=constant\}$. Then the first term in the expression of canonical energy (\ref{waldmass-new}) can be calculated as
\begin{eqnarray}
\int_{\infty}Q[\xi]&=&-\frac{1}{16\pi}\int_{\infty}
\epsilon_{abcd}\nabla^c\xi^d\nonumber\\
&=&-\frac{1}{16\pi}\int_{\infty}\sqrt{-g}\big(
g^{0e}\Gamma^1_{\ e0}
-g^{1e}\Gamma^0_{\ e0}\big)(\d\theta\wedge\d\varphi)_{ab}.
\end{eqnarray}
Direct calculation shows
\begin{eqnarray}
&&g^{0e}\Gamma^1_{\ e0}=-\frac{m}{r^2}+\cdots,\\
&&g^{1e}\Gamma^0_{\ e0}=\frac{16m+\partial_u(C_{AB}C^{AB})}{16r^2}+\cdots,
\end{eqnarray}
and
\begin{eqnarray}
\sqrt{-g}=r^2\sin\theta+\cdots.
\end{eqnarray}
Therefore,
\be
Q[\xi]=\frac{1}{8\pi}\int_{S^2}
\bigg[m+\frac{1}{32}\partial_u\big(C_{AB}C^{AB}\big)\bigg]\sin\theta\d\theta\d\varphi.
\ee

Now we try to express the term $\int_{\infty}\xi^a\Theta_{abc}$ at null infinity as a total variation term. A straightforward  computation yields

\begin{equation}
\begin{split}
\int_{\infty}\xi^a\Theta_{abc}&=\frac{1}{16\pi}\int_{\infty}(\frac{\partial}{\partial u})^a\epsilon_{dabc}
g^{de}g^{fh}(\nabla_f\delta g_{eh}-\nabla_e\delta g_{fh})\\
&=-\frac{1}{16\pi}\int\bigg[({\rm{I})-(\rm{II})-(\rm{III})+(\rm{IV}})\bigg]\d S,
\end{split}
\end{equation}
where
\begin{eqnarray}
(\rm{I})&:=&g^{1e}g^{fh}
\partial_f\delta g_{eh},\\
(\rm{II})&:=&g^{1e}g^{fh}\Gamma^{a}_{\ fh}\delta g_{ea},\\
(\rm{III})&:=&g^{1e}g^{fh}\partial_e\delta g_{fh},\\
(\rm{IV})&:=&g^{1e}g^{fh}\Gamma^a_{\ eh}\delta g_{fa}.
\end{eqnarray}

Recall that
\begin{eqnarray}
&&g_{11}=g_{1A}=0,\\
&&g_{00}=-1+\frac{2m}{r}+\cdots,\\
&&g_{01}=-1+\frac{1}{16r^2}C_{AB}C^{AB}+\cdots,\\
&&g_{0A}=D^EC_{AE}+\bigg[\frac{4}{3}N_A-\frac{1}{8}D_A(C_{EF}C^{EF})\bigg]\frac{1}{r}+\cdots,\\
&&g_{AB}=r^2\bigg(q_{AB}+\frac{C_{AB}}{r}+\cdots\bigg),
\end{eqnarray}
 we proceed to compute the terms
  $(\rm{I}),(\rm{II}),(\rm{III})$ and $(\rm{IV})$ respectively.


Direct calculation shows
\begin{equation}
{(\rm{I})}=\partial_r\delta g_{uu}+\partial_u\delta g_{ur}-
\frac{1}{r^2}\partial_{\theta}\delta g_{u\theta}
-\frac{1}{r^2\sin^2\theta}\partial_{\varphi} \delta g_{u\varphi}+o(\frac{1}{r^2}),
\end{equation}

\begin{equation}
{(\rm{II})}=-\frac{2}{r}\delta g_{uu}+\frac{\cot\theta}{r^2}\delta g_{u\theta}+O(\frac{1}{r^3}),
\end{equation}
\begin{equation}
\begin{split}
(\rm{III})&=2\partial_u\delta g_{ur}-g^{AB}\partial_u\delta g_{AB}-g^{AB}\partial_r \delta g_{AB}+O(\frac{1}{r^3})\\
&=2\partial_u\delta g_{ur}-\frac{1}{r^2}\partial_u\delta g_{\theta\theta}
-\frac{1}{r^2\sin^2\theta}\partial_u\delta g_{\varphi\varphi}\\
&
+\frac{1}{r^2}\partial_r\delta g_{\theta\theta}
+\frac{1}{r^2\sin^2\theta}\partial_r\delta g_{\varphi\varphi}+O(\frac{1}{r^3}),
\end{split}
\end{equation}
\begin{equation}
{(\rm{IV})}=\frac{1}{r^3}\delta g_{\theta\theta}+\frac{1}{r^3\sin^2\theta}\delta g_{\varphi\varphi}
+O(\frac{1}{r^3}).
\end{equation}

From the above results, we find
\begin{equation}\label{C11}
\begin{split}
\int_{\infty}\xi^a\Theta_{abc}&=
-\frac{1}{16\pi}\int_{\infty}\d S\bigg(
(\rm{I})-(\rm{II})-(\rm{III})+(\rm{IV})\bigg)\\
&=-\frac{1}{16\pi}\int_{\infty}\d S\bigg(
\partial_r\delta g_{uu}-\partial_u\delta g_{ur}-\frac{1}{r^2}
\partial_{\theta}\delta g_{u\theta}
-\frac{1}{r^2\sin^2\theta}\partial_{\varphi}
\delta g_{u\varphi}\\
&\ \ \ \ \ \ \ \ +\frac{2}{r}\delta g_{uu}-\frac{\cot\theta}{r^2}\delta g_{u\theta}
+\frac{1}{r^2}\partial_u\delta g_{\theta\theta}
+\frac{1}{r^2\sin^2\theta}\partial_u\delta g_{\varphi\varphi}\\
&\ \ \ \ \ \ \ \  -\frac{1}{r^2}\partial_r\delta g_{\theta\theta}
-\frac{1}{r^2\sin^2\theta}\partial_r\delta g_{\varphi\varphi}+\frac{1}{r^3}\delta g_{\theta\theta}+\frac{1}{r^3\sin^2\theta}\delta g_{\varphi\varphi}\bigg)\\
&=\delta\bigg[-\frac{1}{16\pi}\int_{\infty}\d S\bigg(\partial_r\hat{g}_{uu}-\partial_u\hat{g}_{ur}-\frac{1}{r^2}\partial_{\theta}\hat{g}_{u\theta}\\
&\ \ \ \ \ \ \ \ \  \ \ \ \ \ -\frac{1}{r^2\sin^2\theta}\partial_{\varphi}\hat{g}_{u\varphi}+\frac{2}{r}\hat{g}_{uu}-\frac{\cot\theta}{r^2}\hat{g}_{u\theta}\bigg)\bigg].
\end{split}
\end{equation}
Here $\hat{g}_{\alpha\beta}=(g-g^{M})_{\alpha\beta}$, i.e., the metric tensor subtracting the Minkowski data. For the last equality, we have used the tracefree property of $C_{AB}$,
\begin{eqnarray}
q^{AB}C_{AB}=C_{\theta\theta}+\frac{1}{\sin^2\theta}
C_{\varphi\varphi}=0,
\end{eqnarray}
which is preserved under variation.

Note that
\begin{equation}
\begin{split}
D^A\hat{g}_{uA}&=q^{AB}D_B\hat{g}_{uA}=g^{AB}(\partial_B\hat{g}_{uA}-\Gamma^C_{AB}\hat{g}_{uC})\\
&=\partial_{\theta}\hat{g}_{u\theta}+\frac{1}{\sin^2\theta}\partial_{\varphi}\hat{g}_{u\varphi}+\cot\theta\hat{g}_{u\theta},
\end{split}
\end{equation}
and  we can simplify the expression of $\int_{\infty}\xi^a\Theta_{abc}$ as
\begin{eqnarray}
\int_{\infty}\xi^a\Theta_{abc}=\delta\bigg[-\frac{1}{16\pi}\int_{\infty}\d S\bigg(\partial_r\hat{g}_{uu}-\partial_u\hat{g}_{ur}+\frac{2}{r}\hat{g}_{uu}\bigg)\bigg].
\end{eqnarray}
It follows from
\begin{eqnarray}
\hat{g}_{uu}=\frac{2m}{r},\ \ \hat{g}_{ur}=\frac{1}{16r^2}C_{AB}C^{AB},
\end{eqnarray}
that
\begin{eqnarray}
\int_{\infty}\xi^a\Theta_{abc}&=&\delta\bigg[-\frac{1}{16\pi}\int_{\infty}\d S\bigg(-\frac{2m}{r^2}-\frac{1}{16r^2}\partial_u(C_{AB}C^{AB}))+\frac{4m}{r^2}\bigg]\nonumber\\
&=&\delta\bigg[-\frac{1}{16\pi}\int_{S^2}\bigg(2m-\frac{1}{16}\partial_u(C_{AB}C^{AB})\bigg)\sin\theta\d\theta\d\varphi\bigg].
\end{eqnarray}

Finally, the canonical energy is
\begin{eqnarray}
\mathcal{E}&=&\int_{\infty} Q-\xi\cdot B\nonumber\\
&=&\int_{S^2}\bigg[\frac{1}{16\pi}\bigg(2m+\frac{1}{16}\partial_u(C_{AB}C^{AB})\bigg)\nonumber\\
&&+\frac{1}{16\pi}\bigg(2m-\frac{1}{16}\partial_u(C_{AB}C^{AB})\bigg)\bigg]\sin\theta\d\theta\d\varphi\nonumber\\
&=&\frac{1}{4\pi}\int_{S^2}m\sin\theta\d\theta\d\varphi
\end{eqnarray}
with the 3-form
\begin{equation}\nonumber
B=-\frac{1}{16\pi}\bigg(2m-\frac{1}{16}\partial_u(C_{AB}C^{AB})\bigg)\sin\theta \d u\wedge \d \theta \wedge \d \varphi.\end{equation}

In this section, we have reexamined the Bondi mass at null infinity via the Iyer-Wald formula within the Bondi-Sachs coordinates. Our analysis demonstrates that the Iyer-Wald formula naturally leads to the Bondi mass expression directly in the physical spacetime, thereby bypassing the need of conformal compactification.



\section{Bondi mass of polyhomogeneous spacetimes }\label{s5}

 For smooth asymptotically flat spacetimes, we have successfully employed the Iyer-Wald framework to obtain the charge associated with the asymptotic time translation at null infinity  in the previous section. In 1995, Chru\'{s}ciel, MacCallum and Singleton discussed the Bondi mass of a class of polyhomogeneous spacetimes with $N_3=-\infty$ \cite{Chrusciel95}. Recently, the BMS charges in special polyhomogeneous spacetimes are discussed by using the Barnich-Brandt prescription \cite{Godazagar2020}. In this section, we attempt to derive the mass expression of polyhomogeneous spacetimes by using the Iyer-Wald formula.



Recall that the Iyer-Wald canonical energy is given by \cite[Eq. (83)] {Iyer94}

\be\label{waldmass}
\mathcal{E}\equiv-\Big(\int_{\infty}Q_{bc}-\xi^a B_{abc}\Big).
\ee
Comparing with the original definition in \cite{Iyer94}, there appears a total minus in the expression of the canonical energy. This is because we take the Landau-Lifshitz timelike convention for the signature of the spacetime metric, $(+,-,-,-)$.

For polyhomogeneous spacetimes,  let $\xi^a$ be the asymptotic time translation
\be
\xi^a=(\frac{\partial}{\partial u})^a.
\ee
The 2-sphere at null infinity is referred to as the limit as $r\rightarrow\infty$ of the coordinate spheres $\{r=\mbox{constant},u=\mbox{constant}\}$. Then the first term in the expression of canonical energy (\ref{waldmass}) can be calculated as
\be
\begin{split}
-\int_{\infty}Q[\xi]&=\frac{1}{16\pi}\int_{\infty}
\epsilon_{abcd}\nabla^c\xi^d\\
&=\frac{1}{16\pi}\int_{\infty}\sqrt{-g}\big(
g^{0e}\Gamma^1_{\ e0}
-g^{1e}\Gamma^0_{\ e0}\big)(\d\theta\wedge\d\varphi)_{ab}.
\end{split}
\ee
Direct calculation shows
\be
\begin{split}
g^{0e}\Gamma^1_{\ e0}&=\frac{U_1-U_1'}{r^2}+\cdots,\\
g^{1e}\Gamma^0_{\ e0}&=-\frac{U_1-U_1'}{r^2}+\cdots,
\end{split}
\ee
and
\be
\sqrt{-g}=r^2\sin\theta+\cdots.
\ee
Therefore,
\be
-Q[\xi]=\frac{1}{8\pi}\lim_{r\rightarrow\infty}\int_{S^2}
\big[U_{1}-U_1'\big]\sin\theta\d\theta\d\varphi.
\ee
When $N_3=-\infty$,  we have $U_1=U_{1,0}(\theta,\varphi)$ and
\be
-Q[\xi]=\frac{1}{8\pi}\lim_{r\rightarrow\infty}\int_{S^2}
U_{1,0}\sin\theta\d\theta\d\varphi.
\ee

We now compute the contribution to the canonical energy (\ref{waldmass}) from the second term. Using the symplectic potential 3-form (\ref{potential3form}), we have
\be
\begin{split}
\int_{\infty}\xi\cdot\Theta&=\int_{\infty}(\frac{\partial}{\partial u})^a\Theta_{abc}\\
&=\frac{1}{16\pi}\int_{\infty}
(\frac{\partial}{\partial u})^a\epsilon_{dabc}
g^{de}g^{fh}(\nabla_f\delta g_{eh}-\nabla_e\delta g_{fh})\\
&=-\frac{1}{16\pi}\int_{\infty}
\d S\ g^{1e}g^{fh}\bigg(
\partial_f\delta g_{eh}
-\Gamma^{a}_{\ fh}\delta g_{ea}
-\partial_e\delta g_{fh}
+\Gamma^a_{\ eh}\delta g_{fa}
\bigg)\\
\end{split}
\ee

Let
\be
\begin{split}
(\rm{I}):=&g^{1e}g^{fh}
\partial_f\delta g_{eh},\\
(\rm{II}):=&g^{1e}g^{fh}\Gamma^{a}_{\ fh}\delta g_{ea},\\
(\rm{III}):=&g^{1e}g^{fh}\partial_e\delta g_{fh},\\
(\rm{IV}):=&g^{1e}g^{fh}\Gamma^a_{\ eh}\delta g_{fa}.
\end{split}
\ee
Note that the variation of metric components has the following behaviour
\be\label{C2}
\begin{split}
& \delta g_{uu}\sim O\big(r^{-1}(\ln r)^{N_3+1}\big),\\
& \delta g_{ru}=\delta g_{rr}=\delta g_{r\theta}=\delta g_{r\varphi}=0,\\
& \delta g_{u\theta},\delta g_{u\varphi}\sim O\big((\ln r)^{N_3+1}\big),\\
& \delta g_{\theta\theta},\delta g_{\theta\varphi},\delta g_{\varphi\varphi}\sim O\big(r(\ln r)^{N_3+1}\big).
\end{split}
\ee
Straightforward calculation shows
\be
\begin{split}
(\rm{I})=&\partial_r\delta g_{uu}-
\frac{1}{r^2}\partial_{\theta}\delta g_{u\theta}
-\frac{1}{r^2\sin^2\theta}\partial_{\varphi}\delta g_{u\varphi}+O\big(\frac{(\ln r)^{2N_3+2}}{r^3}\big),\\
(\rm{II})=&-\frac{2}{r}\delta g_{uu}+\frac{\cot\theta}{r^2}\delta g_{u\theta}+O\big(\frac{(\ln r)^{2N_3+2}}{r^3}\big),\\
(\rm{III})=&-\frac{1}{r^2}\partial_u\delta g_{\theta\theta}
-\frac{1}{r^2\sin^2\theta}\partial_u\delta g_{\varphi\varphi}\\
&+\frac{1}{r^2}\partial_r\delta g_{\theta\theta}
+\frac{1}{r^2\sin^2\theta}\partial_r\delta g_{\varphi\varphi}
+O\big(\frac{(\ln r)^{2N_3+2}}{r^3}\big),\\
(\rm{IV})&=\frac{1}{r^3}\delta g_{\theta\theta}+\frac{1}{r^3\sin^2\theta}\delta g_{\varphi\varphi}+O\big(\frac{(\ln r)^{2N_3+2}}{r^3}\big).
\end{split}
\ee
From the above results, we find
\be\label{C11}
\begin{split}
\int_{\infty}\xi^a\Theta_{abc}&=
-\frac{1}{16\pi}\int_{\infty}\d S\bigg(
(\rm{I})-(\rm{II})-(\rm{III})+(\rm{IV})\bigg)\\\
&=-\frac{1}{16\pi}\int_{\infty}\d S\bigg(
\partial_r\delta g_{uu}-\frac{1}{r^2}
\partial_{\theta}\delta g_{u\theta}
-\frac{1}{r^2\sin^2\theta}\partial_{\varphi}
\delta g_{u\varphi}\\
&\ \ \ \ \ \   +\frac{2}{r}\delta g_{uu}-\frac{\cot\theta}{r^2}\delta g_{u\theta}
+\frac{1}{r^2}\partial_u\delta g_{\theta\theta}
+\frac{1}{r^2\sin^2\theta}\partial_u\delta g_{\varphi\varphi}\\
&\ \ \ \ \ \  -\frac{1}{r^2}\partial_r\delta g_{\theta\theta}
-\frac{1}{r^2\sin^2\theta}\partial_r\delta g_{\varphi\varphi}+\frac{1}{r^3}\delta g_{\theta\theta}+\frac{1}{r^3\sin^2\theta}\delta g_{\varphi\varphi}\bigg).
\end{split}
\ee
Note that
\be
\begin{split}
\delta g_{\theta\theta}=&\sqrt{2}r\ \delta(\xi^{\theta}_2+\bar{\xi}^{\theta}_2),\\
\delta g_{\varphi\varphi}=&-i\sqrt{2}r\sin^3\theta\
\delta(\xi^{\varphi}_2-\bar{\xi}^{\varphi}_2)
=-\sqrt{2}r\sin^2\theta\ \delta(\xi^{\theta}_2+\bar{\xi}^{\theta}_2),
\end{split}
\ee
where we have used (\ref{xithetaphi}),
and one arrives at
\be
\begin{split}
\int_{\infty}\xi^a\Theta_{abc}&=-\frac{1}{16\pi}\int_{\infty}\d S\bigg(
\partial_r\delta g_{uu}-\frac{1}{r^2}
\partial_{\theta}\delta g_{u\theta}
-\frac{1}{r^2\sin^2\theta}\partial_{\varphi}
\delta g_{u\varphi}\\
&\ \ \ \ \ \   +\frac{2}{r}\delta g_{uu}-\frac{\cot\theta}{r^2}\delta g_{u\theta}-\frac{1}{r^2}\partial_r\delta g_{\theta\theta}
\\
&\ \ \ \ \ \
-\frac{1}{r^2\sin^2\theta}\partial_r\delta g_{\varphi\varphi}+\frac{1}{r^3}\delta g_{\theta\theta}+\frac{1}{r^3\sin^2\theta}\delta g_{\varphi\varphi}\bigg)\\
&=\delta\bigg{\{}-\frac{1}{16\pi}\int_{\infty}\d S\bigg(
\partial_r\hat{ g}_{uu}-\frac{1}{r^2}
\partial_{\theta}\hat{ g}_{u\theta}
-\frac{1}{r^2\sin^2\theta}\partial_{\varphi}
\hat{ g}_{u\varphi}\\
&\ \ \ \ \ \   +\frac{2}{r}\hat{ g}_{uu}-\frac{\cot\theta}{r^2}\hat{ g}_{u\theta}-\frac{1}{r^2}\partial_r\hat{ g}_{\theta\theta}
\\
&\ \ \ \ \ \
-\frac{1}{r^2\sin^2\theta}\partial_r\hat{ g}_{\varphi\varphi}+\frac{1}{r^3}\hat{ g}_{\theta\theta}+\frac{1}{r^3\sin^2\theta}\hat{ g}_{\varphi\varphi}\bigg)\bigg{\}}
\end{split}
\ee
where
\be\label{ghat}
\begin{split}
\hat{g}_{uu}=&-\frac{2U_1}{r},\\
\hat{g}_{u\theta}=&X_2^{\theta}-\frac{\omega_1+\bar{\omega}_1}{\sqrt{2}},\\
\hat{g}_{u\varphi}=&X_2^{\varphi}\sin^2\theta+\frac{i(\omega_1-\bar{\omega}_1)}{\sqrt{2}}\sin\theta,\\
\hat{ g}_{\theta\theta}=&\sqrt{2}r(\xi^{\theta}_2+\bar{\xi}^{\theta}_2),\\
\hat{g}_{\varphi\varphi}=&-i\sqrt{2}r
(\xi^{\varphi}_2-\bar{\xi}^{\varphi}_2)\sin^3\theta.
\end{split}
\ee

Recall that (\ref{xithetaphi}) gives $\xi_2^{\varphi}=-\frac{i}{\sin\theta}\xi_2^{\theta}$ and hence $\hat{g}_{\varphi\varphi}=-\hat{g}_{\theta\theta}\sin^2\theta,$ which yields
\be
\begin{split}
&\frac{1}{r^2}\partial_u \hat{g}_{\theta\theta}
+\frac{1}{r^2\sin^2\theta}\partial_u \hat{g}_{\varphi\varphi}=0,\\
&\frac{1}{r^2}\partial_r \hat{g}_{\theta\theta}
+\frac{1}{r^2\sin^2\theta}\partial_r \hat{g}_{\varphi\varphi}=0,\\
&\frac{1}{r^3} \hat{g}_{\theta\theta}+
\frac{1}{r^3\sin^2\theta} \hat{g}_{\varphi\varphi}=0.
\end{split}
\ee

Finally, the canonical energy $\mathcal{E}$ defined at null infinity is
\be\label{NUmass}
\begin{split}
\mathcal{E}&=\frac{1}{16\pi}\int_{\infty}\d S\ \bigg(
\frac{2U_1-2U_1'}{r^2}-\partial_r \hat{g}_{uu}+\frac{1}{r^2}
\partial_{\theta} \hat{g}_{u\theta}
+\frac{1}{r^2\sin^2\theta}\partial_{\varphi}
 \hat{g}_{u\varphi}\\
&\ \ \ \ \ \ \ \ \ \ \ \ \ \ -\frac{2}{r} \hat{g}_{uu}+\frac{\cot\theta}{r^2} \hat{g}_{u\theta}
-\frac{1}{r^2}\partial_u \hat{g}_{\theta\theta}
-\frac{1}{r^2\sin^2\theta}\partial_u \hat{g}_{\varphi\varphi}\\
&\ \ \ \ \ \ \ \ \ \ \ \ \ \ +\frac{1}{r^2}\partial_r \hat{g}_{\theta\theta}
+\frac{1}{r^2\sin^2\theta}\partial_r \hat{g}_{\varphi\varphi}-\frac{1}{r^3} \hat{g}_{\theta\theta}-\frac{1}{r^3\sin^2\theta} \hat{g}_{\varphi\varphi}\bigg)\\
&=\frac{1}{16\pi}\int_{\infty}\d S\ \bigg(
\frac{2U_1-2U_1'}{r^2}-\partial_r \hat{g}_{uu}+\frac{1}{r^2}
\partial_{\theta} \hat{g}_{u\theta}\\
&\ \ \ \ \ \ \ \ \ \ \ \ \ \ +\frac{1}{r^2\sin^2\theta}\partial_{\varphi}
 \hat{g}_{u\varphi}-\frac{2}{r} \hat{g}_{uu}+\frac{\cot\theta}{r^2} \hat{g}_{u\theta}\bigg)\\
&=\frac{1}{16\pi}\int_{\infty}\d S\ \bigg(
\frac{2U_1-2U_1'}{r^2}-\frac{2U_1-2U_1'}{r^2}+\frac{1}{r^2}
\partial_{\theta} \hat{g}_{u\theta}
\\
&\ \ \ \ \ \ \ \ \ \ \ \ \ \ +\frac{1}{r^2\sin^2\theta}\partial_{\varphi}
 \hat{g}_{u\varphi}+\frac{4U_1}{r^2}+\frac{\cot\theta}{r^2} \hat{g}_{u\theta}\bigg)\\
&=\frac{1}{16\pi}\int_{\infty}\d S\ \bigg(\frac{4U_1}{r^2}+
\frac{1}{r^2}
\partial_{\theta} \hat{g}_{u\theta}+
\frac{1}{r^2\sin^2\theta}\partial_{\varphi}
 \hat{g}_{u\varphi}+\frac{\cot\theta}{r^2} \hat{g}_{u\theta}\bigg).
\end{split}
\ee
Using (\ref{ghat}), the above expression of the canonical energy $\mathcal{E}$ can be simplified as
\be
\begin{split}\label{NUmass-final}
\mathcal{E}&=\frac{1}{16}\int_{\infty}\d S\bigg[\frac{4U_1}{r^2}+\frac{1}{r^2}(\partial_{\theta} X_2^{\theta}+\partial_{\varphi}X_2^{\varphi}+X_2^{\theta}\cot\theta)\\
&\ \ \ \ \ \ \ \ \ \ \ \ \  -\frac{1}{\sqrt{2}r^2}\bigg(\partial_{\theta}\omega_1-\frac{i}{\sin\theta}\partial_{\varphi}\omega_1+\omega_1\cot\theta\bigg)\\
&\ \ \ \ \ \ \ \ \ \ \ \ \  -\frac{1}{\sqrt{2}r^2}\bigg(\partial_{\theta}\bar\omega_1+\frac{i}{\sin\theta}\partial_{\varphi}\bar\omega_1+\bar\omega_1\cot\theta\bigg)\bigg]\\
&=\frac{1}{16}\int_{\infty}\d S\bigg[\frac{4U_1}{r^2}-\frac{1}{\sqrt{2}r^2}(\bar\eth\omega_1+\eth\bar\omega)\bigg]\\
&=\frac{1}{4\pi}\int_{\infty}\frac{U_1}{r^2}\d S.
\end{split}
\ee

Some remarks are  in order concerning the result of (\ref{NUmass-final}).

(i) When $\Psi_0^3=0$ (but $\Psi_0^4$ need not to be zero),  we have
\be
U_1=-\frac{1}{2}(\Psi_2^{3,0}+\bar\Psi_2^{3,0}).
\ee
Therefore, the canonical energy (\ref{NUmass-final}) becomes
\be
\mathcal{E}=-\frac{1}{8\pi}\int_{S^2}(\Psi_2^{3,0}+\bar\Psi_2^{3,0})\sin\theta\d \theta\d\varphi.
\ee
This is just the `mass' introduced by Newman and Unti in \cite{NU62}. Moreover, recall that
\be
\dot\Psi_2^{3,0}=-\eth^2\dot{\bar\sigma}_{2,0}-\sigma_{2,0}\ddot{\bar\sigma}_{2,0},
\ee
and hence
\be
\begin{split}
\frac{\d \mathcal{E}}{\d u}&=\frac{1}{8\pi}\int_{S^2}\bigg(\eth^2\dot{\bar\sigma}_{2,0}+\sigma_{2,0}\ddot{\bar\sigma}_{2,0}+\bar\eth^2\dot{\sigma}_{2,0}+\bar\sigma_{2,0}\ddot{\sigma}_{2,0}\bigg)\sin\theta\d \theta\d\varphi\\
&=\frac{1}{8\pi}\int_{S^2}\bigg(\sigma_{2,0}\ddot{\bar\sigma}_{2,0}+\bar\sigma_{2,0}\ddot{\sigma}_{2,0}\bigg)\sin\theta\d \theta\d\varphi\\
&=\frac{1}{8\pi}\int_{S^2}\bigg[\frac{\d}{\d u}(\sigma_{2,0}\dot{\bar\sigma}_{2,0}+\bar\sigma_{2,0}\dot{\sigma}_{2,0})-2|\dot\sigma_{2,0}|^2\bigg]\sin\theta\d \theta\d\varphi.\\
\end{split}
\ee
Define the `reduced energy' as
\be
\begin{split}
\hat{\mathcal{E}}&\equiv\mathcal{E}-\frac{1}{8\pi}\int_{S^2}\bigg(\sigma_{2,0}\dot{\bar\sigma}_{2,0}+\bar\sigma_{2,0}\dot{\sigma}_{2,0}\bigg)\sin\theta\d \theta\d\varphi\\
&=-\frac{1}{8\pi}\int_{S^2}\bigg(\Psi_2^{3,0}+\bar\Psi_2^{3,0}+\sigma_{2,0}\dot{\bar\sigma}_{2,0}+\bar\sigma_{2,0}\dot{\sigma}_{2,0}\bigg)\sin\theta\d \theta\d\varphi.
\end{split}
\ee
It possesses a nice monotonicity formula,
\be
\frac{\d \hat{\mathcal{E}}}{\d u}=-\frac{1}{4\pi}\int_{S^2}|\dot\sigma_{2,0}|^2\ \sin\theta\d \theta\d\varphi\leq 0.
\ee
In fact, $\hat{\mathcal{E}}$ is just the well-known Bondi mass. It turns out that this charge coincides with the one obtained by Godazgar and Macaulay in \cite[Eq.(5.10)]{Godazagar2020}, where the BMS charges for polyhomogeneous spacetimes with $\Psi_0^3=0$ are derived by using the Barnich-Brandt prescription \cite{Barnich2002}.

(ii) When $N_3=0$, i.e.,
\be
\Psi_0=\frac{\Psi_0^{3,0}(u,\theta,\varphi)}{r^3}+\frac{\Psi_0^4}{r^4}+\cdots,\ ^{\#}\Psi_0^4=N_4,\cdots.
\ee
Eq.(\ref{2.42}) yields
\be\label{4.31}
\begin{split}
U_1&=U_{1,0}+U_{1,1}z,\ z=\ln r,\\
U_{1,1}&=\gamma_{2,1}+\bar\gamma_{2,1},\\
U_{1,0}&=U_{1,1}+\gamma_{2,0}+\bar\gamma_{2,0}.
\end{split}
\ee
It follows from (\ref{2.40}) that
\be\label{4.32}
\begin{split}
\gamma_{2,1}&=-\frac{1}{2}\Psi_2^{3,1},\\
\gamma_{2,0}&=\frac{1}{2}\bigg[\gamma_{2,1}-\alpha_{1,0}\tau_{2,0}-\beta_1\bar{\tau}_{2,0}-\Psi_2^{3,0}\bigg].
\end{split}
\ee
Eqn.(\ref{2.48}) yields
\be\label{4.33}
\begin{split}
\Psi_2^{3,1}=\bar\eth\Psi_1^{3,0}-\dot{\bar\sigma}_{2,0}\Psi_0^{3,0}
\end{split}
\ee
and (\ref{2.47}) yields
\be\label{4.34}
\begin{split}
\Psi_1^{3,0}=\bar\eth\Psi_0^{3,0}.
\end{split}
\ee
It can be deduced from (\ref{4.31})-(\ref{4.34}) that
\be
\begin{split}
U_1=&\bigg[-\frac{3}{4}(\bar{\eth}^2\Psi_0^3+\eth^2\bar{\Psi}_0^3-\dot{\bar{\sigma}}_{2,0}\Psi_0^3-\dot{\sigma}_{2,0}\bar{\Psi}_0^3)-\frac{1}{2}(\Psi_2^{3,0}+\bar\Psi_2^{3,0})\bigg]\\
&-\frac{1}{2}(\bar{\eth}^2\Psi_0^3+\eth^2\bar{\Psi}_0^3-\dot{\bar{\sigma}}_{2,0}\Psi_0^3-\dot{\sigma}_{2,0}\bar{\Psi}_0^3)z.
\end{split}
\ee

Therefore, (\ref{NUmass-final}) becomes
\be\label{N3=0mass}
\begin{split}
\mathcal{E}=&\frac{1}{4\pi}\int_{S^2}\bigg[\frac{3}{4}(\dot{\bar{\sigma}}_{2,0}\Psi_0^3+\dot{\sigma}_{2,0}\bar{\Psi}_0^3)-\frac{1}{2}(\Psi_2^{3,0}+\bar\Psi_2^{3,0})\bigg]\sin\theta\d \theta\d\varphi\\
&+\frac{1}{8\pi}\int_{S^2}(\dot{\bar{\sigma}}_{2,0}\Psi_0^3+\dot{\sigma}_{2,0}\bar{\Psi}_0^3)z\ \sin\theta\d \theta\d\varphi.
\end{split}
\ee
In general, the quantity $\mathcal{E}$ will blow-up since $z=\ln r \rightarrow \infty$ as $r\rightarrow \infty$. If one assumes that
\begin{equation}
\int_{S^2}\dot{\bar{\sigma}}_{2,0}\Psi_0^3+\dot{\sigma}_{2,0}\bar{\Psi}_0^3=0,
\end{equation}
then the above $\mathcal{E}$ remains finite.

Moreover, regarding the evolution of $\mathcal{E}$ in (\ref{N3=0mass}), we have
\be
\begin{split}
\frac{\d\mathcal{E}}{\d u}&=\frac{1}{4\pi}\int_{S^2}\bigg[\frac{3}{4}(\ddot{\bar\sigma}_{2,0}\Psi_0^{3,0}+\dot{\bar\sigma}_{2,0}\dot\Psi_0^{3,0}+\ddot{\sigma}_{2,0}\bar\Psi_0^{3,0}+\dot{\sigma}_{2,0}\dot{\bar\Psi}_0^{3,0})\\
&\ \ \ \ \ \ \ \ \  \ \ \ -\frac{1}{2}(\dot\Psi_2^{3,0}+\dot{\bar\Psi}_2^{3,0})\bigg]\sin\theta\d \theta\d\varphi\\
&\ \ +\frac{1}{8\pi}\int_{S^2}\bigg[\ddot{\bar\sigma}_{2,0}\Psi_0^{3,0}+\dot{\bar\sigma}_{2,0}\dot\Psi_0^{3,0}+\ddot{\sigma}_{2,0}\bar\Psi_0^{3,0}+\dot{\sigma}_{2,0}\dot{\bar\Psi}_0^{3,0}\bigg]z\ \sin\theta\d \theta\d\varphi.
\end{split}
\ee
Recall that Eqs.(\ref{2.46}) and (\ref{2.48}) give
\be
\dot\Psi_0^{3,0}=0,\ \ \dot\Psi_2^{3,0}=-\eth^2\dot{\bar{\sigma}}_{2,0}-\sigma_{2,0}\ddot{\bar\sigma}_{2,0}.
\ee
Therefore,
\be
\begin{split}
\frac{\d\mathcal{E}}{\d u}&=\frac{1}{4\pi}\int_{S^2}\bigg[\frac{3}{4}(\ddot{\bar\sigma}_{2,0}\Psi_0^{3,0}+\ddot{\sigma}_{2,0}\bar\Psi_0^{3,0})+\frac{1}{2}(\sigma_{2,0}\ddot{\bar\sigma}_{2,0}+\bar\sigma_{2,0}\ddot{\sigma}_{2,0})\bigg]\sin\theta\d \theta\d\varphi\\
&\ \ +\frac{1}{8\pi}\int_{S^2}\bigg[\ddot{\bar\sigma}_{2,0}\Psi_0^{3,0}+\ddot{\sigma}_{2,0}\bar\Psi_0^{3,0}\bigg]z\ \sin\theta\d \theta\d\varphi\\
&=\frac{1}{16\pi}\int_{S^2}\bigg[\partial_u\bigg(3(\dot{\bar\sigma}_{2,0}\Psi_0^{3,0}+\dot{\sigma}_{2,0}\bar\Psi_0^{3,0})+2(\sigma_{2,0}\dot{\bar\sigma}_{2,0}+\bar\sigma_{2,0}\dot{\sigma}_{2,0})\bigg)\\
&\ \ \ \ \ -4|\dot\sigma_{2,0}|^2\bigg]\sin\theta\d \theta\d\varphi+\frac{1}{8\pi}\int_{S^2}\partial_u\bigg(\dot{\bar\sigma}_{2,0}\Psi_0^{3,0}+\dot{\sigma}_{2,0}\bar\Psi_0^{3,0}\bigg)z\ \sin\theta\d \theta\d\varphi.
\end{split}
\ee
Define the `reduced energy' as
\be
\begin{split}
\hat{\mathcal{E}}&\equiv\mathcal{E}-\frac{1}{16\pi}\int_{S^2}\bigg(3(\dot{\bar\sigma}_{2,0}\Psi_0^{3,0}+\dot{\sigma}_{2,0}\bar\Psi_0^{3,0})+2(\sigma_{2,0}\dot{\bar\sigma}_{2,0}+\bar\sigma_{2,0}\dot{\sigma}_{2,0})\bigg)\sin\theta\d \theta\d\varphi\\
&\ \  \ \ \ \ -\frac{1}{8\pi}\int_{S^2}\bigg(\dot{\bar\sigma}_{2,0}\Psi_0^{3,0}+\dot{\sigma}_{2,0}\bar\Psi_0^{3,0}\bigg)z\ \sin\theta\d \theta\d\varphi.
\end{split}
\ee
Then we have
\be
\frac{\d \hat{\mathcal{E}}}{\d u}=-\frac{1}{4\pi}\int_{S^2}|\dot\sigma_{2,0}|^2\ \sin\theta\d \theta\d\varphi\leq 0.
\ee

In summary, we have obtained the explicit expression for the canonical energy, given in (5.18), for vacuum polyhomogeneous spacetimes by applying the Iyer-Wald formalism within Newman-Unti coordinates. When $\Psi_0^3=0$, regardless of whether $\Psi_0^4$ vanishes, this canonical energy reduces to the Newman-Unti mass. By subtracting a suitable quantity, we find that the resulting `reduced energy' (5.23) is just the well-known Bondi mass, which decreases monotonically in the presence of gravitational radiation. Furthermore, for the case $N_3=0$, we define  the `reduced energy' by (5.36), which also possesses a monotonicity property, as shown in (5.37).

\section{Gravitational wave memory and balance law}\label{s6}

The linear memory of gravitational waves (GW) arising from the change in the quadrupole moment of a gravitational wave source was  identified by Zel'dovich {\it et al.} \cite{Zel74, Payne1983, BG1985, BT1987}. Subsequently, Christodoulou and Frauendiener discovered that gravitational waves themselves can generate  memory known as the nonlinear memory \cite{Christo91, Frau92}. The effect of this kind of memory  is related to the relative displacement of nearby observers, so it is also called the displacement memory.   Over the past few years, numerous works have been established on the potential for detecting the nonlinear memory \cite{Seto2009, PBP2010, HL2010, CJ2012, MCC2014, LTLBC2016, MTL2017, DTLW2020}. However, gravitational wave  memory measured by the detector is notably weak, as GW memory behaves mainly as a quasi-direct current signal \cite{Cao21-2}. Consequently, the development of an accurate theoretical model is crucial for the detection of GW memory.
Noticing that the BMS theory does not necessitate the slow and weak field conditions for the GW sources,  Nichols {\it et al.} proposed a method to calculate the GW memory based on the BMS theory \cite{Nichols17, Khera21}. More recently, combining the results of numerical relativity, Cao et al. applied the BMS method to develop a surrogate model that correlates the parameters of binary black holes with GW memory \cite{Cao21-2}. Utilizing this surrogate model, they estimated the GW memory associated with all 48 binary black hole events documented in GWTC-2. In the context of the BMS method for calculating the GW memory, a critical step involves the derivation of the balance law of smooth asymptotically flat spacetimes. For polyhomogeneous spacetimes, it is worthwhile reexamining the balance law, as the asymptotic behaviors of Newman-Penrose (NP) quantities are significantly different between polyhomogeneous spacetimes and smooth asymptotically flat spacetimes.

In Section \ref{s2}, the asymptotic behaviour of NP quantities of polyhomogeneous spacetime have been obtained. It should be emphasized that, albeit $\Psi_0^3$ includes the logarithmic terms, there appear no logarithmic terms in $\Psi_4^1$,
\be
\Psi_4^1=-\ddot{\bar\sigma}_{2,0}.
\ee
Since $\Psi_4^1$ is closely related to  geodesic deviation equation,  from the viewpoint of the detection of gravitational waves, $\Psi_4^1$ or $\sigma_{2,0}$ is the most significant quantity \cite{Mag08}. In fact, $\sigma_{2,0}$ is related to the plus mode $h_+$ and the cross mode $h_{\times}$ of gravitational waves by
\be
\sigma_{2,0}=\frac{D_s}{2}(h_+ +i h_{\times}),
\ee
where $D_s$ is the luminosity distance between the observer and the source \cite{Cao21-1}.

 From the asymptotic behaviour of the NP quantities, we have
\be
\begin{split}
\frac{\partial}{\partial u}(\Psi_2^3+\sigma_2\dot{\bar\sigma}_2)&=
\dot\Psi_2^3+\dot\sigma_2\dot{\bar\sigma}_2+\sigma_2
\ddot{\bar\sigma}_{2}\\
&=
\eth\Psi_3^2+\sigma_2\Psi_4^1+\dot\sigma_{2,0}
\dot{\bar\sigma}_{2,0}+\sigma_2
\ddot{\bar\sigma}_{2,0}\\
&=-\eth^2\dot{\bar\sigma}_{2,0}+
\sigma_2(-\ddot{\bar\sigma}_{2,0})+\dot\sigma_{2,0}
\dot{\bar\sigma}_{2,0}+\sigma_2
\ddot{\bar\sigma}_{2,0}\\
&=|\dot\sigma_{2,0}|^2-\eth^2\dot{\bar\sigma}_{2,0}.
\end{split}
\ee
Consequently,
\be\label{balance1}
\begin{split}
\int_{u_1}^{u_2}\big(|\dot\sigma_{2,0}|^2-
\eth^2\dot{\bar\sigma}_{2,0}\big)\d u&=\big(\Psi_2^3+\sigma_2\dot{\bar\sigma}_2\big)
\bigg|_{u_1}^{u_2}\\
&=\big(\Psi_2^{3,0}+\sigma_{2,0}\dot{\bar\sigma}_{2,0}\big)
\bigg|_{u_1}^{u_2}.
\end{split}
\ee
This is the so-called balance law. Taking the limits $u_1\rightarrow -\infty$ and $u_2\rightarrow\infty$ in (\ref{balance1}), it shows that
\be\label{balance12}
\eth^2\big(\bar{\sigma}_{2,0}\big|_{-\infty}^{\infty}\big)
=-\bigg(\Psi_2^{3,0}+\sigma_{2,0}\dot{\bar\sigma}_{2,0}\bigg)\bigg|_{-\infty}^{\infty}+\int_{-\infty}^{\infty}|\dot\sigma_{2,0}|^2\d u.
\ee
This formula describes the gravitational waves memory in polyhomogeneous spacetimes.
It turns out that \eqref{balance1} and \eqref{balance12} remain unchanged as the ones obtained in smooth asymptotically flat spacetimes \cite{Cao21-1} and \cite{Frau92}.

Pasterski, Strominger and Zhiboedov proposed a new type of gravitational wave memory named the spin memory \cite{Pasterski2016}. It was shown that the effect of the spin memory is the relative time delay of different light rays at very large radial distance $r_0$. Mao and Wu found a new observational effect by looking at timelike geodesics near null infinity \cite{Mao-Wu-2019}. In the rest of this section, we extend Mao and Wu's results to polyhomogeneous asymptotically flat spacetimes.

From Section \ref{s2}, we can obtain the covariant metric components of a polyhomogeneous asymptotically flat spacetime.
Now we consider a timelike hypersurface $\Sigma_{r_0}=\{r=r_0\}$ in a polyhomogeneous spacetime with $r_0$ a large constant. Up to relevant orders, the induced metric $h_{ij}$ reads
\begin{eqnarray}
\d s_{\Sigma}^2&=&h_{ij}\d x^i\d x^j\nonumber\\
&=&\bigg[1-\frac{2U_1}{r_0}+\frac{1}{r_0^2}\bigg(-2U_2-(X_2^{\theta})^2+\sqrt{2}X_2^{\theta}(\omega_1+\bar\omega_1)\nonumber\\
&&-i\sqrt{2}X_2^{\varphi}(\omega_1-\bar\omega_1)\sin\theta-(X_2^{\varphi})^2\sin^2\theta\bigg)+\cdots\bigg]\d u^2\nonumber\\
&&+2\bigg[\bigg(X_2^{\theta}-\frac{\omega_1+\bar\omega_1}{\sqrt{2}}\bigg)+\frac{1}{2r_0}\bigg(2X_3^{\theta}-2\sqrt{2}X_2^{\theta}(\xi_2^{\theta}+\bar\xi_2^{\theta})\nonumber\\
&&+(\xi_2^{\theta}+\bar\xi_2^{\theta})(\omega_1+\bar\omega_1)-\sqrt{2}(\omega_2+\bar\omega_2)+i\sqrt{2}X_2^{\varphi}(\xi_2^{\theta}-\bar\xi_2^{\theta})\sin\theta\nonumber\\
&&-i(\xi_2^{\varphi}+\bar\xi_2^{\varphi})(\omega_1-\bar\omega_1)\sin\theta
-\sqrt{2}X_2^{\varphi}(\xi_2^{\varphi}+\bar\xi_2^{\varphi})\sin^2\theta\bigg)+\cdots\bigg]\d u\d\theta\nonumber\\
&&+2\bigg[\bigg(X_2^{\varphi}\sin^2\theta+\frac{i\sin\theta}{\sqrt{2}}(\omega_1-\bar\omega_1)\bigg)-\frac{i\sin\theta}{2r_0}\bigg(2iX_3^{\varphi}\sin\theta\nonumber\\
&&+\sqrt{2}X_2^{\varphi}(\bar\xi_2^{\varphi}-\xi_2^{\varphi})
-\sqrt{2}X_2^{\theta}(\xi_2^{\theta}-\bar\xi_2^{\theta})-\sqrt{2}(\omega_2-\bar\omega_2)\nonumber\\
&&+(\xi_2^{\theta}-\bar\xi_2^{\theta})(\omega_1+\bar\omega_1)+\sqrt{2}X_2^{\varphi}(\xi_2^{\varphi}-\bar\xi_2^{\varphi})\cos2\theta\nonumber\\
&&-\sqrt{2}iX_2^{\theta}(\xi_2^{\varphi}+\bar\xi_2^{\varphi})\sin\theta-i(\xi_2^{\varphi}-\bar\xi_2^{\varphi})(\omega_1-\bar\omega_1)\sin\theta\bigg)+\cdots\bigg]\d u\d\varphi\nonumber
\end{eqnarray}
\begin{eqnarray}
&&+\bigg[-r_0^2+\sqrt{2}(\xi_2^{\theta}+\bar\xi_2^{\theta})r_0+\frac{1}{2}\bigg(-3(\xi_2^{\theta}+\bar\xi_2^{\theta})^2+2\sqrt{2}(\xi_3^{\theta}+\bar\xi_3^{\theta})\nonumber\\
&&+2i(\xi^{\varphi}_2+\bar\xi_2^{\varphi})(\xi_2^{\theta}-\bar\xi_2^{\theta})\sin\theta-(\xi_2^{\varphi}+\bar\xi_2^{\varphi})^2\sin^2\theta\bigg)+\cdots\bigg]\d \theta^2\nonumber\\
&&+\bigg[\frac{r_0\sin\theta}{\sqrt{2}}\bigg((\xi_2^{\varphi}+\bar\xi_2^{\varphi})\sin\theta-i(\xi_2^{\theta}-\bar\xi_2^{\theta})\bigg)-\frac{i\sin\theta}{2}\bigg(-(\xi_2^{\varphi})^2+(\bar\xi_2^{\varphi})^2\nonumber\\
&&-2(\xi_2^{\theta})^2+2(\bar\xi_2^{\theta})^2+\sqrt{2}(\xi_3^{\theta}-\bar\xi_3^{\theta})+((\xi_2^{\varphi})^2-(\bar\xi_2^{\varphi})^2)\cos2\theta\nonumber\\
&&-2i\sin\theta(\bar\xi_2^{\varphi}\xi_2^{\theta}+\xi_2^{\varphi}\bar\xi_2^{\theta})
+\sqrt{2}i\sin\theta(\xi_3^{\varphi}+\bar\xi_3^{\varphi})\bigg)+\cdots\bigg]\d\theta\d\varphi\nonumber
\end{eqnarray}
\begin{eqnarray}
&&+\bigg[-r_0^2\sin^2\theta-\sqrt{2}ir_0\sin^3\theta(\xi_2^{\varphi}-\bar\xi_2^{\varphi})+\frac{1}{2}\sin^2\theta\bigg((\xi_2^{\theta}-\bar\xi_2^{\theta})^2\nonumber\\
&&+2i\sin\theta(\xi_2^{\varphi}+\bar\xi_2^{\varphi})(\xi_2^{\theta}-\bar\xi_2^{\theta})-2\sqrt{2}i\sin\theta(\xi_3^{\varphi}-\bar\xi_3^{\varphi})\nonumber\\
&&+3(\xi_2^{\varphi}-\bar\xi_2^{\varphi})^2\sin^2\theta\bigg)+\cdots\bigg]\d\varphi^2.
\end{eqnarray}

For free falling observers on the hypersurface $\Sigma_{r_0}$, they move along a timelike geodesic with the tangent vector $V^a$. One thus has
\begin{eqnarray}\label{2.2-new1}
V^iD_iV^j=0,
\end{eqnarray}
where $D_i$ is the covariant derivative operator associated with the induced metric $h_{ij}.$
The asymptotic behaviour of $V^a$ are assumed as \cite{Mao-Wu-2019}
\begin{eqnarray}
    &&V^{u}=1+\sum_{k=1}^{\infty}\frac{V_k^u(u,z_0,\theta,\varphi)}{r_0^k},\\
      &&V^{\theta}=\sum_{k=1}^{\infty}\frac{V_k^{\theta}(u,z_0,\theta,\varphi)}{r_0^k},\\
      && V^{\phi}=\sum_{k=1}^{\infty}\frac{V_k^{\phi}(u,z_0,\theta,\varphi)}{r_0^k},
\end{eqnarray}
where $z_0=\ln r_0.$

The geodesic equation (\ref{2.2-new1}) can be solved order by order.

 At order of $O(\frac{1}{r_0})$, we have
\begin{eqnarray}
\frac{\partial V_1^{\theta}(u,z_0,\theta,\varphi)}{\partial u}=0\ \Rightarrow\ V_1^{\theta}=0.
\end{eqnarray}
\begin{eqnarray}
\frac{\partial V_1^{\varphi}(u,z_0,\theta,\varphi)}{\partial u}=0\ \Rightarrow\ V_1^{\varphi}=0.
\end{eqnarray}
\begin{eqnarray}
&&\frac{\partial V_1^u(u,z_0,\theta,\varphi)}{\partial u}-\frac{\partial U_1(u,z_0,\theta,\varphi)}{\partial u}=0\nonumber\\
&&\Rightarrow V_1^u(u,z_0,\theta,\varphi)=U_1(u,z_0,\theta,\varphi).
\end{eqnarray}
To get the above three equation, we have set the integration constants to be zero since the observers are initially static.

 At order of $O(\frac{1}{r_0^2})$, we have

\begin{eqnarray}
&&\frac{1}{\sqrt{2}}\frac{\partial}{\partial u}(\omega_1+\bar\omega_1)+\frac{\partial V_2^{\theta}}{\partial u}-\frac{\partial X_2^{\theta}}{\partial u}=0\nonumber\\
&&\Rightarrow V_2^{\theta}=X_2^{\theta}-\frac{1}{\sqrt{2}}(\omega_1+\bar\omega_1).
\end{eqnarray}
\begin{eqnarray}
&&-\frac{i}{\sqrt{2}\sin\theta}\frac{\partial}{\partial u}(\omega_1-\bar\omega_1)+\frac{\partial V_2^{\varphi}}{\partial u}-\frac{\partial X_2^{\varphi}}{\partial u}=0\nonumber\\
&&\Rightarrow V_2^{\varphi}=X_2^{\varphi}+\frac{i}{\sqrt{2}\sin\theta}(\omega_1-\bar\omega_1).\label{V2theta-new}
\end{eqnarray}
\begin{eqnarray}
&&\bar\omega_1\frac{\partial\omega_1}{\partial u}+\omega_1\frac{\partial\bar\omega_1}{\partial u}-3U_1\frac{\partial U_1}{\partial u}-\frac{\partial U_2}{\partial u}+
\frac{\partial V_2^u}{\partial u}=0\nonumber\\
&&\Rightarrow\ V_2^u=\frac{3}{2}(U_1)^2+U_2-\omega_1\bar\omega_1.\label{V2phi-new}
\end{eqnarray}

 At order of $O(\frac{1}{r_0^3})$, we have
\begin{eqnarray}
&&-\frac{\partial U_1}{\partial\theta}+\bar\xi_2^{\theta}\frac{\partial\omega_1}{\partial u}+\xi_2^{\theta}\frac{\partial\bar\omega_1}{\partial u}+\frac{1}{\sqrt{2}}
\frac{\partial\bar\omega_2}{\partial u}+\frac{1}{\sqrt{2}}\frac{\partial\bar\omega_2}{\partial u}+\frac{\partial V_3^{\theta}}{\partial u}\nonumber\\
&&+\frac{1}{\sqrt{2}}\frac{\partial U_1}{\partial u}(\omega_1+\bar\omega_1)-X_2^{\theta}\frac{\partial U_1}{\partial u}-\frac{\partial X_2^{\theta}}{\partial u}U_1-\frac{\partial X_3^{\theta}}{\partial u}\nonumber\\
&&+\frac{U_1}{\sqrt{2}}\frac{\partial(\omega_1+\bar\omega_1)}{\partial u}+
\bar\omega_1\frac{\partial\xi_2^{\theta}}{\partial u}+\omega_1\frac{\partial\bar\xi_2^{\theta}}{\partial u}=0\nonumber\\
&&\Rightarrow V_3^{\theta}=-\frac{1}{\sqrt{2}}U_1(\omega_1+\bar\omega_1)+X_2^{\theta}U_1
+X_3^{\theta}-\frac{1}{\sqrt{2}}(\omega_2+\bar\omega_2)\nonumber\\
&&\ \ \ \ \  \ \ \ \ \ \ -\xi_2^{\theta}\bar\omega_1-\bar\xi_2^{\theta}\omega_1+\int^u\frac{\partial U_1(v,z_0,\theta,\varphi)}{\partial\theta}\d v.
\end{eqnarray}
\begin{eqnarray}
&&-\frac{1}{\sin^2\theta}\frac{\partial U_1}{\partial\varphi}+\bar\xi_2^{\varphi}\frac{\partial\omega_1}{\partial u}
+\xi_2^{\varphi}\frac{\partial\bar\omega_1}{\partial u}-\frac{i}{\sqrt{2}\sin\theta}
\frac{\partial\omega_2}{\partial u}+\frac{i}{\sqrt{2}\sin\theta}
\frac{\partial\bar\omega_2}{\partial u}\nonumber\\
&&-\frac{i(\omega_1-\bar\omega_1)}{\sqrt{2}\sin\theta}\frac{\partial U_1}{\partial u}-X_2^{\varphi}\frac{\partial U_1}{\partial u}-U_1\frac{\partial X_2^{\varphi}}{\partial u}+\frac{\partial V_3^{\varphi}}{\partial u}-\frac{\partial X_3^{\varphi}}{\partial u}\nonumber\\
&&+\frac{iU_1}{\sqrt{2}\sin\theta}\frac{\partial(\bar\omega_1-\omega_1)}{\partial u}+\bar\omega_1\frac{\partial\xi_2^{\varphi}}{\partial u}+\omega_1\frac{\partial\bar\xi_2^{\varphi}}{\partial u}=0\nonumber\\
&&\Rightarrow V_3^{\varphi}=\frac{i}{\sqrt{2}\sin\theta}(\omega_2-\bar\omega_2)+X_2^{\varphi}U_1+X_3^{\varphi}+\frac{iU_1}{\sqrt{2}\sin\theta}(\omega_1-\bar\omega_1)\nonumber\\
&&\ \ \ \ \ \ \ \  \ \ \ \ -\xi_2^{\varphi}\bar\omega_1-\bar\xi_2^{\varphi}\omega_1+\frac{1}{\sin^2\theta}\int^u\frac{\partial U_1(v,z_0,\theta,\varphi)}{\partial\varphi}\d v.
\end{eqnarray}

At order of $r_0^{-2}$, it follows from $(\ref{V2theta-new})$ and $(\ref{V2phi-new})$ that $V^a$ has angular components. The infinitesimal change of the proper time $\chi$ of the observer can be calculated as follows. Direct calculation shows the covector of the observer is
\begin{eqnarray}
\d\chi&=&h_{ij}V^i\d x^j\nonumber\\
&=&\d u-\frac{1}{r_0}\bigg[U_1\d u+\bigg(\int^u\frac{\partial U_1}{\partial\theta}\d v\bigg)\d\theta
+\bigg(\int^u\frac{\partial U_1}{\partial\varphi}\d v\bigg)\d\varphi\bigg]+O(\frac{1}{r_0^2})\nonumber\\
&=&\d\bigg(u+\frac{1}{r_0}\mathcal{M}\bigg)+O(\frac{1}{r_0^2}),
\end{eqnarray}
where
\begin{eqnarray}
\mathcal{M}:=-\int^uU_1(v,z_0,\theta,\varphi)\d v.
\end{eqnarray}
The change of the proper time $\Delta\chi$ between two spacetime points $(u_i,\theta_i,\varphi_i)$ and $(u_f,\theta_f,\varphi_f)$ on the geodesic reads
\begin{eqnarray}
\Delta\chi=\Delta u+\frac{1}{r_0}\Delta\mathcal{M}+O(\frac{1}{r_0^2}).
\end{eqnarray}
This means the observer receives a time delay of order $O(\frac{1}{r_0})$. It should be emphasized that $\Delta\mathcal{M}$ is angle dependent.

To see the effect of the logarithmic term, we consider the special case with
\begin{eqnarray}
&&\Psi_0^3=\Psi_0^{3,0}.
\end{eqnarray}
For this case, it follows from Section \ref{s2} that
\begin{eqnarray}
&&U_1=U_{1,0}+U_{1,1}z_0,\\
&&U_{1,1}=\gamma_{2,1}+\bar\gamma_{2,1},\\
&&\gamma_{2,1}=-\frac{1}{2}\bar\eth\bar\eth\Psi_0^{3,0}+\frac{1}{2}\dot{\bar{\sigma}}_{2,0}\Psi_0^{3,0},\\
&&U_{1,0}=\frac{3}{2}(\gamma_{2,1}+\bar\gamma_{2,1})-\frac{1}{2}(\Psi_2^{3,0}+\bar\Psi_2^{3,0}).
\end{eqnarray}
The Bianchi identity gives
\begin{eqnarray}
\frac{\partial\Psi_1^{4,0}}{\partial u}=\eth\Psi_2^{3,0}-(\eth\dot{\bar{\sigma}}_{2,0})(2\sigma_{2,0}-\Psi_0^{3,0}),
\end{eqnarray}
which in turn yields
\begin{eqnarray}\label{bs2}
\bar\eth\eth\Delta\mathcal{M}&=&\Delta\bigg[
\frac{1}{2}\bar\eth\Psi_1^{4,0}+\frac{1}{2}\eth\bar\Psi_1^{4,0}
+(\bar\eth\sigma_{2,0})(\eth\bar\sigma_{2,0})\bigg]\nonumber\\
&+&\int_{u_i}^{u_f}\bigg[\sigma_{2,0}
(\bar\eth\eth\dot{\bar{\sigma}}_{2,0})+
\bar\sigma_{2,0}(\eth\bar\eth\dot{{\sigma}}_{2,0})\bigg]\d v\nonumber\\
&+&\int_{u_i}^{u_f}\bigg[\frac{1}{2}
\bar\eth(\Psi_0^{3,0}\eth\dot{\bar{\sigma}}_{2,0})+
\frac{1}{2}\eth(\bar\Psi_0^{3,0}\bar\eth\dot{{\sigma}}_{2,0})
+\frac{3}{2}\bar\eth\eth(\gamma_{2,1}+\bar\gamma_{2,1})\bigg]\d v\nonumber\\
&+&\bigg[\int_{u_i}^{u_f}\bar\eth\eth(\gamma_{2,1}+
\bar\gamma_{2,1})\d v\bigg]z_0.
\end{eqnarray}
When $\Psi_0^{3,0}=0$,  the above equation reduces to
\begin{eqnarray}
\bar\eth\eth\Delta\mathcal{M}&=&\Delta\bigg[\frac{1}{2}
\bar\eth\Psi_1^{4,0}+\frac{1}{2}\eth\bar\Psi_1^{4,0}
+(\bar\eth\sigma_{2,0})(\eth\bar\sigma_{2,0})\bigg]\nonumber\\
&+&\int_{u_i}^{u_f}\bigg[\sigma_{2,0}
(\bar\eth\eth\dot{\bar{\sigma}}_{2,0})
+\bar\sigma_{2,0}(\eth\bar\eth\dot{{\sigma}}_{2,0})\bigg]\d v.
\end{eqnarray}
This is exactly the same as Eq.(4.8) in \cite{Mao-Wu-2019}.

As shown in \cite{Mao-Wu-2019}, by choosing a ring of freely falling observers who are initially static and synchronised, the difference in the proper time between each observer on this ring at a later time encodes the gravitational wave-form, providing an observable signature of spin memory.

\section{Conclusions}\label{s7}
In this article, we investigate the mass and the memory effect of polyhomogeneous spacetimes. The gravitational waves persist all the way to the null infinity. The analysis of the solutions of the Einstein field equations was studied in the basic settings by Bondi {\it et al.} \cite{Bondi62, Sachs62, NP62} using the formal series expansions in powers of $1/r$. Carving out the feature of null geometry, Penrose \cite{P63} proposed the idea of conformal compactifications. It requires that the conformal structure should be smoothly extended aross the null infinity. The implicit assumption of the smoothness of the coformal extension leads to the very specific fall-off behaviours of isolated gravitating
 bodies and the emitting radiation. This is knowns as the  peeling-off property. If there is a slower decay assumption for the component of the Weyl tensors, then the logarithmic terms will appear in the asymptotic expansions. It seems that, regardless of the geometric elegance, the conformal compactification is a guess proposal and 
the purely $1/r$ expansions are too stringent. One kind of spacetimes removing the smoothness assumption are those that admit a polyhomogeneous expansion. Precisely, the metrics can be expanded in terms of a combination of powers of $1/r$ and $\ln r$.

 The asymptotic behaviour of the Newman-Penrose quantities of polyhomogeneous spacetime, including the metric coefficients, the spin coefficients, and the tetrad components of the Weyl tensor are obtained in vacuum. This generalization is nontrivial since the calculational complexity substantially increases when considering polyhomogeneous expansions. For smooth asympotically flat spacetimes, we have reexamined the Bondi mass at null infinity via the Iyer-Wald formula within the Bondi-Sachs coordinates. Our analysis demonstrates that the Iyer-Wald formula naturally leads to the Bondi mass expression directly in the physical spacetime, avoiding the difficulties and subtle isuues arising from the (unphysical) conformal compactification \cite{Grant2022}.

For polyhomogeneous spacetimes, we calculate the charge associated with the asymptotic translation at null infinity via the Iyer-Wald formalism and obtain the canonical energy $\mathcal{E}$ within the Newman-Unti gauge. The canonical energy $\mathcal{E}$ reduces to the Newman-Unti mass when the logarithmic terms vanish. However, the canonical energy does not has the monotonically decreasing property with respect to the retraded time. To find an energy expression which admits a mass-loss formula, we propose a new notion of energy $\hat{\mathcal{E}}$, called the `reduced energy'. This gives a possible way to define the Bondi mass of polyhomogeneous spacetimes. In particular, for polyhomogeneous spacetimes with $\Psi_0^3=0$, the result is consistent with the one obtained by Godazgar and Macaulay in \cite[Eq.(5.10)]{Godazagar2020} using the Barnich-Brandt prescription \cite{Barnich2002}. It should also be remarked that our method here differs from the previous work in \cite{Wald2000, Grant2022} and the gauge choice we adopt is slightly different from the one used in \cite{Kroon99}. 

Gravitational wave memory of polyhomogeneous spacetimes are discussed. In conclusion, we find the appearance of the logarithmic terms in the expansion does not change the balance law. This is a mathematical consequence of the symmetries of the Bianchi identities. More precisely, this result relies on the evolution equation for $\Psi_2^3+\sigma_2\bar\sigma_2$ which involves only the leading-order shear $\sigma_2(z)$ and its derivatives. The logarithmic terms in the full expansion of $\sigma_2(z)$  are of higher order in $1/r$ and do not have contributions in the memory effect. It indicates that the gravitational memory seems a good physically measurable effect even for spacetimes with slower decays.

It has to be mentioned that we suspect that the canonical energy via the Iyer-Wald formula ceases to be finite for general polyhomogeneous spacetimes. Under certain conditions, one is able to define a `reduced energy' satisfying a nice mass loss formula. A deep investigation of the physical interpretation of the `reduced energy' will be the subject of the future work.

\section*{Acknowledgments}
 X. He is partially supported by the Natural Science Foundation of Hunan Province
(2023JJ30179)  and the National Natural Science Foundation of China (12475049). X. Wu was partially supported by the National Natural Science Foundation of China (12275350). N. Xie was partially sponsored by the Natural Science Foundation of Shanghai (24ZR1406000).

\section*{Data availability}
No new data were created or analysed in this study.



\end{document}